\RequirePackage{fixltx2e}
\documentclass[preprint,12pt]{elsarticle}
\usepackage[utf8x]{inputenc}
\usepackage[T1]{fontenc}
\usepackage[english]{babel}
\usepackage{amsmath,amsthm,amssymb,cmll,mathtools}
\usepackage{tikz}
\usepackage{fouriernc} % Changement de police
\usepackage[autostyle = true, english = american, strict = true, autopunct=true]{csquotes} % guillemets

\usepackage{multirow} 
\usepackage{blkarray} % Pour annoter les matrices
\usepackage{tabularx}
\usepackage{booktabs}
\usepackage{array}

\usepackage{calc} % Calcul de longueurs.

\usepackage{graphicx}
\usepackage{microtype}
\usepackage{hyphenat} % Amélioration des césures, commande \hyp{}

\tikzset{
 >=latex, % option pour de jolies flêches.
 inner sep=0pt,
 outer sep=2pt,
}
\usetikzlibrary{calc} % Calcul de coordonnées

\usepackage{macrosCoNL} % Macros
\usepackage{hyperref}
\usepackage[capitalize%,nameinlink
]{cleveref}

\theoremstyle{plain}
\newtheorem{theorem}{Theorem}
\newtheorem{corollary}[theorem]{Corollary}
\newtheorem{lemma}[theorem]{Lemma}
\newtheorem{proposition}[theorem]{Proposition}

\theoremstyle{definition}
\newtheorem{definition}[theorem]{Definition}

\theoremstyle{remark}
\newtheorem*{remark}{Remark}

\hypersetup{
	pdfdisplaydoctitle		= 	true,			% display document title instead of file name in title bar
	pdfinfo={
		Author	=	{Clément Aubert, Thomas Seiller},
		Title		=	{Logarithmic Space and Permutations},
		Subject	=	{Linear Logic, Complexity, Geometry of Interaction, Pointer Machine, Finite Automata},
		Keywords	=	{Theoretical computer Science, Linear Logic, Implicit complexity, Proof nets, Operator algebra},
		Creator	=	{PDFLaTeX},
		Producer	=	{PDFLaTeX},
		Licence 	= 	{Creative Commons BY-NC-SA 3.0 or later},
		Lang		=	{en}
	}
}

\journal{INFORMATION \& COMPUTATION}

\begin{document}

\begin{frontmatter}

\title{Logarithmic Space and Permutations\tnoteref{t1}}
\tnotetext[t1]{Work partially supported by the ANR projects ANR-10-BLAN-0213 LOGOI and ANR-08-BLAN-0211 COMPLICE.}

\author[lipn]{Clément Aubert}
\ead{aubert@lipn.fr}
\author[ihes]{Thomas Seiller}
\ead{seiller@ihes.fr}
\address[lipn]{LIPN -- UMR CNRS 7030 Institut Galilée - Université Paris-Nord 99, avenue J.-B. Clément 93430 Villetaneuse -- France}
\address[ihes]{Institut des Hautes Études Scientifiques, Le Bois-Marie, 35 Route de Chartres, 91440 Bures-sur-Yvette, France}

\begin{abstract}
In a recent work, Girard proposed a new and innovative approach to computational complexity based on the proofs-as-programs correspondence. In a previous paper, the authors showed how Girard's proposal succeeds in obtaining a new characterization of co-NL languages as a set of operators acting on a Hilbert Space. In this paper, we extend this work by showing that it is also possible to define a set of operators characterizing the class L of logarithmic space languages. 
\end{abstract}


\end{frontmatter}

\section{Introduction}

\subsection{Linear Logic and Implicit Computational Complexity}

Logic, and more precisely proof theory -- the domain whose purpose is the formalization and study of mathematical proofs -- recently yielded numerous developments in theoretical computer science. These developments are founded on a correspondence, often called Curry-Howard correspondence, between mathematical proofs and computer programs (usually formalized in lambda-calculus). The main interest of this correspondence lies in its dynamic nature: program execution corresponds to a procedure on mathematical proofs known as the cut-elimination procedure.

In the eighties, Jean-Yves Girard discovered linear logic through a study of mathematical models of the lambda-calculus. This logical system, as a direct consequence of this correspondence between proofs and programs, is particularly interesting from the point of view of the mathematical foundations of computer science for its resource-awareness. In particular, it gave birth to a number of developments in the field of implicit computational complexity, for instance through the definition and study of restricted logical systems (sub-systems of linear logic) in which the set of representable functions captures a complexity class. For instance, elementary linear logic (ELL) restricts the rules governing the use of exponential connectives -- the connectives dealing with the duplication of the arguments of a function -- and the set of representable functions in ELL is exactly the set of elementary time functions~\cite{danos2003}. It was also shown~\cite{schopp2007} that a characterization of logarithmic space 
computation can be obtained if one restricts both the rules of exponential connectives and the use of universal quantifiers. Finally, a variation on the notion of linear logic \emph{proof nets} succeeded in characterizing the classes \cc{NC} of problems that can be efficiently parallelized~\cite{terui2004}.%,mogbil2009,aubert2011}.

%Linear logic has already given birth to a number of developments in the field of Implicit Computational Complexity. In particular through the definition of constrained logical systems -- obtained by restricting the rules governing the exponential connectives -- which characterize some complexity classes~\cite{danos01}. It was also shown~\cite{schopp07} that a characterization of logarithmic space computation can be obtained if one restricts both the rules of exponential connectives and the use of universal quantifiers. Finally, a variation on the notion of linear logic proof nets succeeded in characterizing the classes \textbf{NC} of problems that can be efficiently parallelized~\cite{terui04, aubert11}.

%Those characterizations were made possible because the Curry-Howard correspondence naturally applies to quantitative considerations.}

\subsection{Geometry of Interaction}
A deep study of the formalization of proofs in linear logic, in particular their formalization as proof nets, led Jean-Yves Girard to initiate a program entitled \emph{geometry of interaction} (GoI)~\cite{girard1989b}. This program, in a first approximation, intends to define a semantics of proofs that accounts for the dynamics of the cut-elimination procedure. Through the correspondence between proofs and programs, this would define a semantics of programs that accounts for the dynamics of their execution. However, the geometry of interaction program is more ambitious: beyond the mere interpretation of proofs, its purpose is to completely reconstruct logic around the dynamics of the cut-elimination procedure. This means reconstructing the logic of programs, where the notion of formula -- or type -- accounts for the behavior of algorithms.

Informally, a geometry of interaction is defined by a set of \emph{paraproofs} together with a notion of interaction, in the same way one defines strategies and their composition in game semantics. An important tool in the construction is a binary function that measures the interaction between two paraproofs. With this function one defines a notion of orthogonality that corresponds to the negation of logic and reconstructs the formulas as sets of paraproofs equal to the orthogonal of a given set of paraproofs: a formula is therefore a set of \enquote{programs} that interact in a similar way to a given set of \emph{tests}.

Since the introduction of this program Jean-Yves Girard proposed different constructions to realize it. These constructions share the notion of paraproofs: operators in a von Neumann algebra. They however differ on the notion of orthogonality they use: in the first constructions, this notion was founded on the nilpotency of the product of two operators, while the more recent construction~\cite{girard2011a} uses Fuglede-Kadison determinant -- a generalization of the usual determinant of matrices that can be defined in type $\textnormal{II}_{1}$ factors.

Since the reconstruction of logic is based on the notion of execution, geometry of interaction constructions are particularly interesting for the study of computational complexity. It is worth noting that the first construction of GoI~\cite{girard1989a} allowed Abadi, Gonthier, and Lévy~\cite{gonthier1992} to explain the optimal reduction of $\lambda$-calculus defined by Lamping~\cite{lamping1989}. This first GoI construction was also used to obtain results in the field of implicit computational complexity~\cite{baillot2001}.

\subsection{A new approach to complexity}

Recently Jean-Yves Girard proposed a new approach for the study of complexity classes that was inspired by his latest construction of a geometry of interaction. Using the crossed product construction between a von Neumann algebra and a group acting on it, he proposed to characterize complexity classes as sets of operators obtained through the internalization of outer automorphisms of the type $\textnormal{II}_{1}$ hyperfinite factor. The authors showed in a recent paper~\cite{aubert2014ctemp} that this approach succeeds in defining a characterization of the set of \cc{co-NL} languages as a set of operators in the type $\textnormal{II}_{1}$ hyperfinite factor. The proof of this result was obtained through the introduction of \emph{non-deterministic pointer machines}, which are abstract machines designed to mimic the computational behavior of operators. The result was obtained by showing that a \cc{co-NL} complete problem could be solved by these machines.

In this paper, we extend these results in two ways. The first important contribution is that we give an alternative proof of the fact that \cc{co-NL} is indeed characterized by non-deterministic pointer machines. This new proof consists in showing that pointer machines can simulate the well-known and studied two-way multi-head finite automata~\cite{hartmanis1972,holzer2008}. The second contribution of this paper consists in obtaining a characterization of the class $\cc{L}$ as a set of operators in the hyperfinite factor of type $\text{II}_{1}$. By studying the set of operators that characterize the class $\cc{co-NL}$ and in particular the encoding of non-deterministic pointer machines as operators, we are able to show that the operators encoding a deterministic machine satisfy a condition expressed in terms of norm. We then manage to show that the language decided by an operator satisfying this norm condition is in the class \cc{L}, showing that the set of all such 
operators characterizes \cc{L}.

%\note{We proved that those machines were able to decide a \cc{co-NL}-complete set, and we extend this result in this paper by proving that they can simulate $2$-way non-deterministic finite automata, a model that naturally represent logspace computation. Moreover, we define a subset of the operators characterizing \cc{co-NL} using a criteria on the norm, and prove that their computational behavior is equivalent to the one of deterministic pointer machines. By doing so, we manage to prove that \cc{L} is characterized by a set of operators and enhance our proof that this is the case for \cc{co-NL} too.}

\section{The Basic Picture}

The construction uses an operator-theoretic construction known as the crossed product of an algebra by a group acting on it. The interested reader can find a quick overview of the theory of von Neumann algebras in the appendix of the second author's work on geometry of interaction~\cite{seiller2012a}, and a brief presentation of the crossed product construction in the authors' previous work~\cite{aubert2014ctemp} on the characterization of \cc{co-NL}. For a more complete presentation of the theory of operators and the crossed product construction, we refer to the well-known series of Takesaki~\cite{takesaki2001,takesaki2003,takesaki2003a}.

In a nutshell, the crossed product construction $\vn{A}\rtimes_{\alpha} G$ of a von Neumann algebra $\vn{A}$ and a group action $\alpha: G\rightarrow \textnormal{Aut}(\vn{A})$ defines a von Neumann algebra containing $\vn{A}$ and unitaries that internalize the automorphisms $\alpha(g)$ for $g\in G$. For this, one considers the Hilbert space\footnote{The construction $L^{2}(G,\hil{H})$ is a generalization of the well-known construction of the Hilbert space of square-summable functions: in case $G$ is considered with the discrete topology, the elements are functions $f:G\rightarrow \hil{H}$ such that $\sum_{g\in G}\norm{f(g)}^{2}<\infty$.} $\hil{K}=L^{2}(G,\hil{H})$ where $\hil{H}$ is the Hilbert space $\vn{A}$ is acting on, and one defines two families of unitary operators\footnote{Recall that in the algebra $\B{\hil{H}}$ of bounded linear operators on the Hilbert space $\hil{H}$ (we denote by $\inner{\cdot}{\cdot}$ its inner product), there exists an anti-linear involution $(\cdot)^{\ast}$ such that for any 
$\xi,\eta\in\hil{H}$ and $A\in\B{\hil{H}}$, $\inner{A\xi}{\eta}=\inner{\xi}{A^{\ast}\eta}$. This \emph{adjoint operator} coincides with the conjugate-transpose in the algebras of square matrices. A \emph{unitary operator} $u$ is an operator such that $uu^{\ast}=u^{\ast}u=1$.} in $\B{\hil{K}}$:
\begin{itemize}
\item the family $\{\pi(u) \mid u\in\vn{A}\textnormal{ unitary}\}$ which is a representation of $\vn{A}$ on $\hil{K}$;
\item the family $\{\lambda(g) \mid g\in G\}$ which contains unitaries that internalize the automorphisms $\alpha(g)$.
\end{itemize}
Then the crossed product $\vn{A}\rtimes_{\alpha} G$ is the von Neumann algebra generated by these two families.

As a by-product of Girard's work on geometry of interaction~\cite{girard2011a}, we know\footnote{A detailed explanation of this representation can be found in \cref{example}, in the authors' previous work~\cite{aubert2014ctemp} or in the second author's PhD thesis~\cite{seiller2012}.} how to represent integers as operators in the type $\textnormal{II}_{1}$ hyperfinite factor $\finhyp$.

This representation comes from logical considerations, and it has some specificities, among which the fact that the integer is represented as a \emph{circular} binary string, and the fact that every bit is considered as having an input and an output\footnote{Something that will be proven useful in \cref{section-pm-operators}, for it will help us to determine in which direction the integer is read.}.
This representation arises from the interpretation of proofs of the type $\forall X \oc (X\multimap X) \multimap \left( \oc (X\multimap X)\multimap \oc (X\multimap X)\right)$ (the type of binary lists in Elementary Linear Logic) in the setting of geometry of interaction.
% [{\cite{aubert2014ctemp} for a detailed explanation of the representation of integers).
The obtained interpretation is the image of a matrix through an embedding of the matrix algebra in which it is contained in the hyperfinite factor.
Since there are no natural choice of a particular embedding, a given integer does not have a unique representation in $\finhyp$, but a family of representations. Luckily, any two representations of a given integer are unitarily equivalent (we denote by $\vn{M}_{k}(\complexN)$ the algebra of $k\times k$ matrices over $\complexN$):

\begin{proposition}[{{\cite[Proposition 10]{aubert2014ctemp}}}]
Let $N,N'\in\vn{M}_{6}(\complexN)\otimes \finhyp$ be two representations of a given integer. There exists a unitary $u\in\finhyp$ such that:
\[N=(1\otimes u)^{\ast}N'(1\otimes u)\]
\end{proposition}

The next step is then to define a representation of machines, or algorithms, as operators that do not discriminate two distinct representations of a given integer, i.e.\ operators that act uniformly on the set of representations of a given integer. As the authors have shown in their previous work, the crossed product construction allows one to characterize an algebra of operators acting in such a uniform way. We recall the construction in the particular case of the group of finite permutations, which will be the only setting that will be studied in this work.

The algebra $\finhyp$ can be embedded in the infinite tensor product algebra $\vn{T}=\bigotimes_{n\in\naturalN} \finhyp$ through the morphism $\iota_{0}: x\mapsto x\otimes 1\otimes 1\otimes\dots$. We will denote by $\vn{N}_{0}$ the image of $\finhyp$ in $\vn{T}$ through $\iota_{0}$. Notice that this tensor product algebra is isomorphic to $\finhyp$. The idea of Girard is then to use the action of the group $\mathcal{S}$ of finite permutations of $\naturalN$ onto $\vn{T}$, defined by:
\[\sigma.(x_{0}\otimes x_{1}\otimes \dots\otimes x_{k}\otimes \dots)=x_{\sigma^{-1}(0)}\otimes x_{\sigma^{-1}(1)}\otimes \dots\otimes x_{\sigma^{-1}(k)}\otimes\dots\]
This group action defines a sub-algebra $\vn{G}$ of the crossed product algebra $\vn{K}=(\bigotimes_{n\in\naturalN} \finhyp)\rtimes\mathcal{S}$. This algebra $\vn{G}$ is the algebra generated by the family of unitaries $\lambda(\sigma)$ for $\sigma\in\mathcal{S}$, and we think of it as the algebra of machines, or algorithms. As it turns out, the elements of this algebra act uniformly on the set of representations of a given integer:

\begin{proposition}[{{\cite[Proposition 11]{aubert2014ctemp}}}]
Let $N,N'\in\vn{M}_{6}(\complexN)\otimes \vn{N}_{0}$ be two representations of a given integer, and $\phi\in\vn{M}_{6}(\complexN)\otimes\vn{G}$. Then:
\[\phi N\textnormal{ is nilpotent iff }\phi N'\textnormal{ is nilpotent}\]
\end{proposition}

To be a bit more precise, we will be interested in the elements of the algebras of operators $\vn{M}_{6}(\complexN)\otimes\vn{G}\otimes \vn{M}_{k}(\complexN)$. By tensoring with a matrix algebra, which we call the \emph{algebra of states}, we will be able to develop more computational power. An element of one of these algebras will be called an \emph{observation}, and it still acts uniformly on distinct representations $N_{n},N_{n}'$ of a given integer: if $\phi$ is an observation, $\phi (N_{n}\otimes 1_{\vn{M}_{k}(\complexN)})$ is nilpotent if and only if $\phi (N_{n}'\otimes 1_{\vn{M}_{k}(\complexN)})$ is nilpotent\footnote{We will in the following simply write $1$ for the identity element of $\vn{M}_{k}(\complexN)$.}. From this proposition, one can justify that the following definition makes sense:
\begin{definition}
Let $\phi\in\vn{M}_{6}(\complexN)\otimes\vn{G}\otimes \vn{M}_{k}(\complexN)$ be an observation. We define the language accepted by $\phi$ by ($N_{n}$ denotes any representation of the integer $n$):
\[[\phi]=\{n\in\naturalN \mid \phi (N_{n}\otimes 1)\textnormal{ nilpotent}\}\]
By extension, if $P$ is a set of observations, we denote by $[P]$ the set $\{[\phi] \mid \phi\in P\}$.
\end{definition}

To sum up the construction, one has the following objects:
\begin{itemize}
\item an algebra containing representations of integers: the hyperfinite factor $\finhyp$ of type $\textnormal{II}_{1}$, embedded in $\vn{K}$ through the morphism $\pi\circ\iota_{0}$;
\item an algebra containing the representations of machines, or algorithms: the von Neumann sub-algebra $\vn{G}$ of $\vn{K}$ generated by the set of unitaries $\{\lambda(\sigma) \mid \sigma\in\mathcal{S}\}$;
\item a notion of acceptance: if $N$ is a representation of a integer and $\phi$ is a representation of a machine, we say that $\phi$ accepts $N$ when $\phi N$ is nilpotent.
\end{itemize}

\section{Pointer Machines}
We previously introduced~\cite{aubert2014ctemp} non-deterministic pointers machines that were designed to mimic the computational behavior of operators: they do not have the ability to write, their input tape is cyclic, and they are \enquote{universally non-deterministic}, i.e.\ rejection is meaningful whereas acceptation is the default behavior. We exhibited a non-deterministic pointer machine that decides a \cc{co-NL}-complete language and showed how a reduction could be performed with pointers. Here, we complete this result by simulating multi-head finite automata~\cite{rosenberg1966}, a method that allows us to get a deeper understanding of pointer machines, in both their deterministic and non-deterministic versions.

A pointer machine is given by a set of pointers that can move back and forth on the input tape and read (but not write) the values it contains, together with a set of states. For $1 \leqslant i \leqslant p$, given a pointer $p_{i}$, only one of three different \emph{instructions} can be performed at each step: $p_{i}+$, i.e.\ \enquote{move one step forward}, $p_{i}-$, i.e.\ \enquote{move one step backward} and $\epsilon_{i}$, i.e.\ \enquote{do not move}. In the following definition, we let 
%$I_{\{1,\dots,p\}} = \{p_{i}+, p_{i}-, \epsilon_{i} \mid i\in\{1,\dots,p\}\}$
\(I^k= \{p_{k}+, p_{k}-, \epsilon_{k}\}$
be the set of instructions for the \(k\)-th pointer 
and $\Sigma = \{ 0, 1, \star\}$ be the alphabet.
% We will denote by $\# p_i$ the value of the pointer (the address it points at).
We will denote by $\# p_i$ the address of the pointer.

% \begin{definition}[Instructions]
% Let $p_i$ be a pointer. We will denote by $\# p_i$ the value of the pointer (the address it points at). We define the following instructions:
% \begin{itemize}
% \item $p_i+$ which means ``move the $i$-th pointer forward'', i.e.\ let $\#p_i$ be $\#p_i+1$;
% \item $p_i-$ which means ``move the $i$-th pointer backward'', i.e.\ let $\#p_i$ be $\#p_i-1$;
% \item $\epsilon_i$ which means ``do not move the $i$-th pointer''.
% \end{itemize}
% When $p_i$ moves the new value at $\#p_i$ is read and stored. We write $P$ \Note{Plutôt $P_{p}$} the set of instructions $\{p_{i}+, p_{i}-, \epsilon_{i} \mid i\in\{1,\dots,p\}\}$.
% \end{definition}

\begin{definition}
A non-deterministic pointer machine with $p \geqslant 1$ pointers is a couple $M = \{ Q, \rightarrow\}$ where $Q$ is the set of \emph{states} and $\rightarrow \subseteq (\Sigma^p \times Q)\times 
\prod_{i=1}^{p} \left(I^{i}
%\times ((I_{\{1,\dots,p\}}^p
\times (Q \cup \{\textbf{accept, reject}\})\right)$ is the \emph{transition relation} and it is total. We write NDPM($p$) the set of non-deterministic pointer machines with $p$ pointers.
\end{definition}

A configuration of $M \in\text{NDPM}(p)$ is as usual a \enquote{snapshot} of $M$ at a given time, and we define a \emph{pseudo-configuration} $c$ of $M \in \text{NDPM}(p)$ as a \enquote{partial snapshot}: $c \in \Sigma^p \times Q$ contains the last values read by the $p$ pointers and the current state, \emph{but does not contain the addresses of the $p$ pointers}.
It is therefore impossible to resume the computation from a pseudo-configuration provided the description of $M$ and the input, but one can know what would be the set of instructions and the new state resulting from the application of $\to$.
 The set of pseudo-configurations of a machine $M$ is written $C_M$ and it is the domain of the transition relation. If $\rightarrow$ is functional, $M$ is a \emph{deterministic} pointer machine. We write DPM($p$) the set of deterministic pointer machines with $p$ pointers.

Let $M \in$ NDPM($p$), $s \in C_M$ and $n\in \naturalN$ an input. We define $M_{s}(n)$ as $M$ with $n$ encoded as a string\footnote{Of course, one could do the exact same work taking binary words instead of integers. This choice of presentation was originally driven by Girard's will to enlighten that, among the infinitely many isomorphic representations of an integer, \enquote{none of them \textins{was} more \enquote{standard} than the others.}~\cite[p.~244]{girard2012}} on its circular input tape (as $\star a_1 \hdots a_k$ for $a_1 \hdots a_k$ the binary encoding of $n$ and $a_{k+1} = a_0 = \star$) starting in the pseudo-configuration $s$ with $\#p_{i}=0$ for all $1 \leqslant i \leqslant p$ (that is, the pointers are initialized with the address of the symbol $\star$). The pointers may be considered as variables that have been declared but not initialized yet. They are associated with \emph{memory slots} that store the values and are updated only when the pointer moves, so as the pointers did not move yet, those memory slots haven't been initialized. The \emph{initial pseudo-configuration} $s$ initializes those $p$ registers, not necessarily in a faithful way (it may not reflect the values contained at $\#p_i$). The entry $n$ is \emph{accepted} (resp.\ \emph{rejected}) by $M$ with \emph{initial pseudo-configuration} $s \in C_M$ if after a finite number 
of transitions every branch of $M_s(n)$ reaches \textbf{accept} (resp.\ at least a branch of $M$ reaches \textbf{reject}). We say that $M_s(n)$ halts if it accepts or rejects $n$ and that $M$ decides a set $S$ if there exists a pseudo-configuration $s \in C_M$ such that $M_s(n)$ accepts if and only if $n \in S$. 

As we will see, this notion of non-deterministic pointer machines is similar to the classical notion of multi-head finite automata:

% \Note{\begin{definition}[Definition 1\cite{DBLP:journals/corr/abs-0906-3051}]
% For $k \in \naturalN^*$, a non-deterministic two-way $k$-head finite automaton is a tuple $M = \{ S, A, k, \vartriangleright, \vartriangleleft, s_0 , F, \sigma \}$ where $S$ is the finite set of \emph{states}, $A = \{0, 1\}$ is the \emph{alphabet}\footnote{We take it to be equal to $\{0, 1\}$, as any alphabet is equivalent to this one modulo a reasonable translation.}, $k$ is the number of heads, $\vartriangleright$ and $\vartriangleleft$ are the \emph{left and right endmarkers}, $s_0 \in S$ is the \emph{initial state}, $F \subseteq S$ is the set of \emph{accepting states}, and $\sigma \subseteq (S \times (A \cup \{\vartriangleright, \vartriangleleft\})^k) \times (S \times \{−1, 0, 1\}^k)$ is the \emph{transition relation}, where $-1$ means to move the head one square to the left, $0$ means to keep the head on the current square and $1$ means to move it one square to the right. Whenever\footnote{Of course, one may see $\sigma$ as a partial function from $S \times (A \cup \{\vartriangleright, \vartriangleleft\})^k$ to $\wp(S \times \{−1, 0, 1\}^k)$, and that justifies this notation that we will use from now on.} $(s′ , (d_1 , \hdots , d_k )) \in \sigma (s, (a_1 , \hdots , a_k ))$, then $d_i \in \{0, 1\}$ if $a_i = \vartriangleright$, and $d_i \in \{−1, 0\}$ if $a_i = \vartriangleleft$, $1 \leqslant i \leqslant k$. We write $2\text{NDFA}(k)$ the set of non-deterministic two-ways $k$-head finite automata.
% \end{definition}

\begin{definition}[{\cite[Definition 1]{holzer2008}}]
For $k \geqslant 1$, a non-deterministic two-way $k$-head finite automaton is a tuple $M = \{ S, A, k, \vartriangleright, \vartriangleleft, s_0 , F, \sigma \}$ where:
\begin{itemize} 
\item $S$ is the finite set of \emph{states};
\item $A$ is the \emph{alphabet}\footnote{We take it to be always equal to $\{0, 1\}$, as any alphabet is equivalent to this one modulo a reasonable translation.};
\item $k$ is the number of heads;
\item $\vartriangleright$ and $\vartriangleleft$ are the \emph{left and right endmarkers};
\item $s_0 \in S$ is the \emph{initial state};
\item $F \subseteq S$ is the set of \emph{accepting states};
\item $\sigma \subseteq (S \times (A \cup \{\vartriangleright, \vartriangleleft\})^k) \times (S \times \{−1, 0, 1\}^k)$ is the \emph{transition relation}, where $-1$ means to move the head one square to the left, $0$ means to keep the head on the current square and $1$ means to move it one square to the right.
\end{itemize}
Moreover, whenever\footnote{Of course, one may see $\sigma$ as a partial function from $S \times (A \cup \{\vartriangleright, \vartriangleleft\})^k$ to $\wp(S \times \{−1, 0, 1\}^k)$, and that justifies this notation that we will use from now on.} $(s′ , (d_1 , \hdots , d_k )) \in \sigma (s, (a_1 , \hdots , a_k ))$, then $a_i = \vartriangleright$ ($1 \leqslant i \leqslant k$) implies $d_i \in \{0, 1\}$, and $a_i = \vartriangleleft$ ($1 \leqslant i \leqslant k$) implies $d_i \in \{−1, 0\}$. 

In the following, we will denote by $2\text{NDFA}(k)$ the set of non-deterministic two-ways $k$-head finite automata.
\end{definition}

One defines configurations and transitions in a classical way. One should however notice that the heads cannot move beyond the endmarkers because of the definition of the transition relation.

Let $M \in 2\text{NDFA}(k)$ and $n\in \naturalN$ an input whose binary writing is $w$.
We say that $M$ \emph{accepts} $n$ if $M$ starts in state $s_0$, with $\vartriangleright w \vartriangleleft$ written on its input tape and all of its heads on $\vartriangleright$, and if after a finite number of transitions at least one branch of $M$ reaches a state belonging to $F$. We say that $M$ always halt if for all input all branches of $M$ reach after a finite number of transitions a configuration such that no transition may be applied. If $\sigma$ is functional, $M$ is said to be \emph{deterministic}. We write $2\text{DFA}(k)$ the set of deterministic two-way $k$ heads automata.

\section{Simulation and first results}

Multi-head finite automata were introduced in the early seventies~\cite{hartmanis1972} and provided many promising perspectives\footnote{An overview of the main results and open problems of the theory was recently published~\cite{holzer2008}.}. Their nice characterization of \cc{L} and \cc{NL} and the similarities they share with non-deterministic pointer machines will be the key to prove (anew) that non-deterministic pointer machines characterize \cc{co-NL} and that their deterministic restriction characterizes \cc{L}.
%Both are ``pointing devices'', without the ability to write, even when it's called ``read-only head'' in a case and ``pointer'' in the other, it is the same kind of approach that can be reciprocally simulated, as stated in [ref thm].

We will denote by \cc{DPM}, \cc{NDPM}, \cc{$2$DFA} and \cc{$2$NDFA} the sets of languages decided by respectively the sets $\cup_{k \geqslant 1}\textnormal{DPM($k$)}$, $\cup_{k \geqslant 1}\textnormal{NDPM($k$)}$, $\cup_{k \geqslant 1}2\textnormal{DFA($k$)}$ and $\cup_{k \geqslant 1}2\textnormal{NDFA($k$)}$.

\begin{theorem}[{\cite[p.~338]{hartmanis1972}}]\label{FA-et-L}
%%%L’article est un peu chaotique, pas de numéro de théorème précis. Je laisse la page, à voir si on garde ou pas.
 $\cc{L} = \cc{$2$DFA}$ and $\cc{NL} = \cc{$2$NDFA}$.
\end{theorem}

We will denote in the following by $\mathcal{L}(X)$ (resp.\ $\overline{\mathcal{L}(X)}$) the language accepted (resp.\ rejected) by $X$.

\begin{proposition}\label{prophalt}
For all $M \in 2\text{NDFA}(k)$ (resp.\ $M \in 2\text{DFA}(k)$), there exists $k'$ and $M' \in 2\text{NDFA}(k')$ (resp.\ $M' \in 2\text{DFA}(k')$) such that $M'$ always halt and $\mathcal{L}(M) = \mathcal{L}(M')$.
\end{proposition}

\begin{proof}
Given an input $n \in \naturalN$, we know that the number of configurations of $M$ is bounded by $\card{S} \times %(\card{A\cup\{\vartriangleright, \vartriangleleft\}})^k \times
(\log_{2}(n) + 2)^k$, that is to say by $\log_{2}(n)^d$ for $d$ a constant fixed with $M$. We will construct $M'$ so that it halts rejecting after performing more than this number of transitions.

 We set $k' = k + d + 1$, and we construct $M'$ so that its $k$ first heads act as the heads of $M$, and the $d+1$ heads act as the hands of a clock, going back and forth between the two endmarkers. For all $s \in S$ of $M$, $S'$ contains the set of states $s \times \{\rightarrow, \leftarrow\}^{d+1}$ that give the current direction of the $d+1$ heads. At every transition the $k+1$-th head moves according to its direction. For $k < i < k'$, when the $i$-th head has made a round-trip on $n$, the $i+1$-th head moves one square according to its direction. If the $k'$-th head has made a round-trip on $n$ -- that is, is back on $\vartriangleright$ -- $M'$ halts rejecting\footnote{That can be simply done by halting the computation in a state not belonging to $F$.}.

%If $M$ did not stop after $\card{S} \times 2^k \times (\log_{2}(n) + 2)^k$ transitions, it means that it was stuck in a loop and therefore will never accept.
Now remark that if a branch of \(M\) accepts, there exists a branch of \(M\) that accepts without ever looping, i.e.\ without going through the same configuration twice.
Moreover, branches without loops are of length at most
$\card{S} \times (\log_{2}(n) + 2)^k$.
Thus, if $M$ has no branches of length less than $\log_{2}(n)^d$ for the suitable \(d\), then it has no accepting branch.
%So if \(M\) makes more than $\card{S} \times (\log_{2}(n) + 2)^k$ transitions, it will never accept.
By construction, $M'$ always halts after $\log_{2}(n)^{d+1}$ transitions and accepts exactly as $M$ does, so we proved that $\mathcal{L}(M) = \mathcal{L}(M')$ and that $M'$ always halts. Notice that if $M$ was deterministic, so is $M'$.
%, so this result may be restricted to the deterministic case.
\end{proof}

\begin{proposition}\label{prop-sim}
For all $k \geqslant 1$, for all $M \in 2\text{NDFA}(k)$ (resp.\ $M \in 2\text{DFA}(k)$) that always halts, there exists $M' \in \text{NDPM}(k)$ (resp.\ $M' \in 2\text{DFA}(k)$) such that $\overline{\mathcal{L}(M)} =\mathcal{L}(M')$.
\end{proposition}

\begin{proof}
There is only little work to do, since NDPMs are essentially \enquote{re-arrangements} of NDFAs.
Given \(M = \{ S, A, k, \vartriangleright, \vartriangleleft, s_0 , F, \sigma \}\), we design \(M' = \{Q, \to\} \in \text{NDPM}(k)\) that rejects an input iff \(M\) accepts it.
%We suppose without loss of generality that \(A = \{0, 1\}\), and
We set \(Q = S \setminus F\).
The transition relation \(\to\) is obtained from \( \sigma\) as follows:
\begin{itemize}
\item If the state of the resulting configuration of a transition belongs to \(F\), the same transition belongs to $\rightarrow$ with its right-hand side replaced by \textbf{reject};
\item The role of both endmarkers $\vartriangleright$ and $\vartriangleleft$ is played by the symbol $\star$;
\item The instructions $-1, 0$ and $1$ in position $i \in \naturalN$ are translated by $p_i-, \epsilon_i$ and $p_i+$.
\end{itemize}

There is one point that needs some care.
If a transition \(t_1\) in \(\sigma\) have for premise \((q, (\hdots, \vartriangleleft, \hdots))\) and another one \(t_2\) have for premise \((q, (\hdots, \vartriangleright, \hdots))\), we cannot translate in \(\to\) both with \((q, (\hdots, \star, \hdots))\): that would introduce non-determinism in deterministic machines, and could lead to false positive as well as false negative.
Remark that this \enquote{critical pair} may involve more than one pointer.

If such a situation occurs in \(\sigma\), one may patch it by creating duplicates of each states, to label them with the last directions of each pointer.
%\begin{itemize}
%\item One may add extra pointers that \enquote{remember} the directions of the last movements of the pointers involved in the \enquote{conflict} between \(t_1\) and \(t_2\).
%\item One may create duplicates of each states, to label them with the last directions of each pointer.
%One may create duplicates of each states, to label them with the last directions of the pointers involved in the conflict.
%\end{itemize}
%Both those 
This modification provokes a global rewriting of \(\to\) to take into account new states, but do not raise any particular difficulty.
Thanks to a slight overhead in the number of state, it is possible to encode if \(\star\) was read from the right (and hence encode \(\vartriangleright\)) or from the left (and hence encode \(\vartriangleleft\)).

At last, to make $\rightarrow$ total, one just have to complement it by adding, from any pseudo-configuration not in the domain of definition of $\sigma$, transitions leading to \textbf{accept}.
Finally, we chose the initial pseudo-configuration \(c\) to be \(\{\star, \hdots, \star, s_0\}\).

As \(M\) always halts, rejection is exactly \enquote{not accepting}, and $M'$ always halts.
One can check that $M'_{c}(n)$ accepts iff $M(n)$ rejects.
Moreover, if $M$ is deterministic, so is $M'$.
\end{proof}

This proposition has as an immediate corollary the inclusions $\cc{co-$2$DFA} \subseteq \cc{DPM}$ and $\cc{co-$2$NDFA} \subseteq \cc{NDPM}$. By combining this result with \cref{FA-et-L} and the fact that $\cc{L}$ is closed under complementation\footnote{Using the fact that $\cc{NL}=\cc{co-NL}$~\cite{Immerman1988}, we could show that $\cc{NL}\subset\cc{NDPM}$. However, we choose not to use the well-known equality between $\cc{NL}$ and $\cc{co-NL}$ in the hope that this new approach to complexity may provide a new proof of this result. Moreover, it is easier to grasp the computation of observations as \enquote{universally non-deterministic}.}, we obtain the expected result:
%\Note{Of course, for any $M \in 2\text{DFA}(k)$, it is trivial to build $M' \in 2\text{DFA}(k)$ such that $co-\mathcal{L}(M) = \mathcal{L}(M')$, but we had in all generality to prove that $\textbf{co-$2$DFA} \subseteq \cc{DPM}$ and that $\textbf{co-$2$NDFA} \subseteq \cc{NDPM}$ (a corollary of Proposition \ref{prop-sim}), because non-deterministic pointer machines are rather suited to decide complementary languages. (Mal dit : je ne comprends pas ce que tu veux dire à cet endroit)} We may now use Theorem \ref{FA-et-L} to obtain the expected result:

\begin{corollary}
 $ \cc{L} \subseteq \cc{DPM}$ and $\cc{co-NL} \subseteq \cc{NDPM}$.
\end{corollary}

%There exists many variants of the Multi-head finite automata: they can have ``sensing heads'', we may restrict the heads so that they never move to the left (``one-way Finite Automata''), or they can be non-deterministic.
% In the general model of multi-head finite automata the moves of the heads are only depending on the input, and a head cannot feel the presence of another head at the same position at the same time.

The converse inclusion will be a by-product of the encoding of NDPMs into operators and won't be proved directly, but it is reasonable to think of NDPMs as \enquote{modified} $2$NFAs.
Those modifications are introduced to ease the simulation by operators: we already mentioned that the transition relation was total, the input circular, the importance of the initial pseudo-configuration to initialize the computation, but we should also mention that any NDPM can be modified so that it does not move more than one pointer at each transition.
\section{Pointer Machines and Operators}\label{section-pm-operators}

In this section, we will briefly explain the encoding of pointer machines as operators.
The non-deterministic case, which was already defined in our previous work~\cite{aubert2014ctemp}, will be given in \cref{encod-non-det}.
The specifics of the deterministic case, which is a novelty, concludes in \cref{encod-det} this section.
An example, detailed in \cref{example}, will help the reader to grasp the main steps of that encoding.

\subsection{Encoding NDPMs}\label{encod-non-det}

We will begin by a definition of two families of observations.

\begin{definition}
Let $\phi\in\vn{M}_{6}(\complexN)\otimes\vn{G}\otimes\vn{M}_{k}(\complexN)$ be an observation that we will write as a $6k\times 6k$ matrix $(a_{i,j})_{1\leqslant i,j\leqslant 6k}$ with coefficients in $\vn{G}$. We say that $\phi$ is:
\begin{itemize}
\item \emph{A positive observation} if for all $1\leqslant i,j\leqslant 6k$, we have $a_{i,j}=\sum_{m=0}^{l} \alpha^{i,j}_{m}\lambda(g^{i,j}_{m})$ where $l$ is an integer (possibly null), and $\alpha^{i,j}_{m}$ are positive real numbers.
\item \emph{A boolean observation} if for all $1\leqslant i,j\leqslant 6k$, we have $a_{i,j}=\sum_{m=0}^{l} \lambda(g^{i,j}_{m})$ where $l$ is an integer (possibly null).
\end{itemize}
\end{definition}

\begin{definition}
We define the following sets:
\begin{align*}
P_{\geqslant 0} &= \{\phi \mid \phi\textnormal{ is a positive observation}\}\\
P_{+} &= \{\phi \mid \phi\textnormal{ is a boolean observation}\}
\end{align*}
\end{definition}

We showed~\cite{aubert2014ctemp} that any non-deterministic pointer machine $M$ can be encoded as an operator in $P_{+}$, implying that $\cc{co-NL}\subseteq [P_{+}]$.

The encoding of non-deterministic pointer machines was defined as follows: we encode each couple $(c,t)\in\rightarrow$ by an operator $\phi_{c,t}$, and then the encoding of the transition relation corresponds to the sum:
%Given $c \in C_M$ a pseudo-configuration and for all $t$ such that $c \rightarrow t$, we will define $\phi_{c,t}$ an operator that encode the fact that the NDPM makes a transition from the pseudo-configuration $c$. All the possibles transitions from the pseudo-configuration $c$ are then encoded by $\sum_{c \rightarrow t} \phi_{c,t}$, and the transition function is encoded by the operator
\begin{equation*}
\rightarrow^{\bullet}=\sum_{c\in C_M} \left( \sum_{t\text{ s.t. }c \rightarrow t} \phi_{c,t} \right)
\end{equation*}

To define the encoding correctly, we need to introduce a number of additional states. We will denote by $Q^{\uparrow}$ the extended set of states obtained from $Q$.
According to the notation introduced in \cref{operateurs1,operateurs2}, this set has cardinality $\card{Q}+(p\times(\card{Q}^{2}+2))$ and is (a subset of):
\[Q^{\uparrow} = Q \cup \bigcup_{j = 1}^{p} \left(\{ \textbf{mov-back}_{j}, \textbf{back}_{j}\} \cup \{\textbf{mov}^{\textbf{q,q'}}_{j} \mid \textbf{q}, \textbf{q'} \in Q\}\right)\]
The machine will then be encoded as an operator in $\vn{M}_{6}(\vn{G})\otimes \vn{P}$, where $\vn{P}$ is the algebra of \emph{pseudo-states}, i.e.\ an algebra that will encode the memory slots that contain the last value read by the pointers, and the extended set of states $Q^{\uparrow}$. That is:
\[\vn{P}=\underbrace{\vn{M}_{6}(\complexN)\otimes\vn{M}_{6}(\complexN)\otimes \dots\otimes\vn{M}_{6}(\complexN)}_{p\text{ times}}\otimes \vn{M}_{\card{Q^{\uparrow}}}(\complexN)\]
Summing up, the encoding of a non-deterministic pointer machine with a set of states $Q$ will be an observation in the algebra $\vn{M}_{6}(\complexN)\otimes\vn{G}\otimes\vn{M}_{k}(\complexN)$ with  $k=6^{p}\times\card{Q^{\uparrow}}=6^{p}\times \left(\card{Q}+ ( p\times(\card{Q}^{2}+2))\right)$.

The following encoding differs from the one used in our earlier work~\cite{aubert2014ctemp}%ça fait beaucoup de citation du papier, non ?
. The reason for this (small) change is that the encoding of the \enquote{move forward} and \enquote{move backward} instructions were defined as operators whose norm was strictly greater than $1$. Since we want the encoding of a deterministic machine to be of norm at most $1$ [{the next subsection), it is necessary to define a more convenient encoding. The encoding of acceptance and rejection are not modified, but we recall them anyway.
The reader should also keep in mind that the encoding of the integer is somehow redundant, for every bit will have an input and an output: this will help us to discriminate between the movements of a pointer from right to left and the movements from left to right.

Before describing the encoding in details, we need to introduce several notations that will be used for the definition of operators.
We recall that the integer is encoded as a circular string, whose start and end is $\star$, and that every bit has an output that is connected to the input of the following bit.
So for instance \enquote{reading} $0i$
% is wondering
amounts to asking (the integer)
what is the bit before the bit $0$ under treatment, \enquote{reading} $\star o$
%is wondering 
amounts to asking
what is the first bit of the integer.
The way the \enquote{pointer} will parse the entry is highly interactive, or dynamic, for it is always on the edge of reading the next bit.
So we define the projections $\pi_{\ast}$ of $\vn{M}_{6}(\complexN)$ ($\ast\in\{0i,0o,1i,1o,e,s\}$) as the projections on the subspace induced by the basis element\footnote{To be coherent with the notations of our earlier paper, we will always denote by $\{0i,0o,1i,1o,e,s\}$ the basis of $\vn{M}_{6}(\complexN)$ we are working with.}.
We will sometimes denote $e$ (resp.\ $s$) by $\star i$ (resp.\ $\star o$), and we also define the projections $\piin=\pi_{0\textnormal{i}}+\pi_{1\textnormal{i}}+\pi_{e}$ and $\piout=\pi_{0\textnormal{o}}+\pi_{1\textnormal{o}}+\pi_{s}$. Those projections correspond to the following matrices.
\begin{center}
\begin{tabular}{cc}
\(\protect{\piin} =
\begin{pmatrix}
1& 0& 0& 0& 0& 0\\
0& 0& 0& 0& 0& 0\\
0& 0& 1& 0& 0& 0\\
0& 0& 0& 0& 0& 0\\
0& 0& 0& 0& 1& 0\\
0& 0& 0& 0& 0& 0
\end{pmatrix}\)
&
\(\protect{\piout} =
\begin{pmatrix}
0& 0& 0& 0& 0& 0\\ 
0& 1& 0& 0& 0& 0\\
0& 0& 0& 0& 0& 0\\
0& 0& 0& 1& 0& 0\\
0& 0& 0& 0& 0& 0\\
0& 0& 0& 0& 0& 1
\end{pmatrix}\)
\end{tabular}
\end{center}

We will moreover use matrices named $( \text{in} \rightarrow \text{out} )$ and $( \text{out} \rightarrow \text{in} )$, shown below. Those matrices will be used for continuing a movement in the same direction as the last movement, i.e.\ if the integer answered that the last read value was a $0$, it will answer on the basis element $0i$ -- for \enquote{0 in} -- if the pointer is moving forward. Then, to ask the next value, the pointer needs to ask the next value on the basis element $0o$ -- for \enquote{0 out}. To go from $0i$ to $0o$, one can apply the matrix $( \text{in} \rightarrow \text{out} )$. The matrix $( \text{out} \rightarrow \text{in} )$ plays the same role when moving in the opposite direction. Moreover, changes of directions are dealt with the projections $\piin$ and $\piout$.

\begin{center}
\begin{tabular}{cc}
\(\protect{( \text{in} \rightarrow \text{out} )} =
\begin{pmatrix}
0& 0& 0& 0& 0& 0\\
1& 0& 0& 0& 0& 0\\
0& 0& 0& 0& 0& 0\\
0& 0& 1& 0& 0& 0\\
0& 0& 0& 0& 0& 0\\
0& 0& 0& 0& 1& 0
\end{pmatrix}\)
&
\(\protect{( \text{out} \rightarrow \text{in} )} =
\begin{pmatrix}
0& 1& 0& 0& 0& 0\\
0& 0& 0& 0& 0& 0\\
0& 0& 0& 1& 0& 0\\
0& 0& 0& 0& 0& 0\\
0& 0& 0& 0& 0& 1\\
0& 0& 0& 0& 0& 0
\end{pmatrix}\)
\end{tabular}
\end{center}

For the sake of simplicity, we also define the following operators in $\vn{P}$: if $c=(a_{1},\dots,a_{p}, \textbf{q})$ and $c'=(a'_{1},\dots,a'_{p}, \textbf{q}')$, we define the partial isometry:
\begin{equation*}
(c\rightarrow c')=(a_{1}\rightarrow a'_{1})\otimes\dots\otimes(a_{p}\rightarrow a'_{p})\otimes (\textbf{q}\rightarrow \textbf{q}')
\end{equation*}
where $(p\rightarrow p')$ (for $p,p'\in Q^{\uparrow}$ or $p,p'\in\{0i,0o,1i,1o,e,s\}$) is defined as:
\begin{equation*}
\begin{array}{cc}	
	&
	\begin{array}{ccccc}
	~& ~&p&~&~
	\end{array}\\~\\
	\begin{array}{c}
	~\\
	~\\
	p'\\
	~\\
	~
	\end{array}
	&
	\left(\begin{array}{ccccc}
	0 & \dots & 0 & \dots & 0\\
	%\vdots & \vdots & 0 & \vdots & \vdots\\
	\vdots & \ddots & \vdots & \reflectbox{$\ddots$} & \vdots\\
	0 & \dots & 1 & \dots & 0\\
	%\vdots & \vdots & 0 & \vdots & \vdots\\
 \vdots & \reflectbox{$\ddots$} & \vdots & \ddots & \vdots\\
	0 & \dots & 0 & \dots & 0
	\end{array}\right)
	\end{array}
\end{equation*}
For \textbf{S} a set of states, we will use the notation $(\textbf{S} \rightarrow a'_{i})$ (denoted $(\rightarrow a'_{i})$ when $\textbf{S}$ contains all possible states) for the sum of partial isometries $\sum_{s\in\textbf{S}} (s\rightarrow a'_{i})$ that encodes a transition from any state in $\textbf{S}$ to $a'_{i}$. The notation $(\mathbf{S}\rightarrow)$ will denote the projection on the subspace generated by $\mathbf{S}$.
% (si $X$ est l'ensemble des valeurs possibles pour $a_{i}$, c'est un raccourci de notation pour $(X\rightarrow a'_{i})$).

A transition that impacts only on the values stored in the subset of pointers\footnote{In fact this subset will always be a singleton, for our encoding corresponds to a NDPM that moves at most one pointer at each transition and any NDPM can be replaced by another NDPM that moves its pointers one at a time.} $p_{i_{1}},\dots,p_{i_{l}}$ and the state $\textbf{q}$ will be denoted by
\begin{equation*}
((a_{i_{1}}\rightarrow a_{i_{1}}')_{i_{1}};\dots;(a_{i_{l}}\rightarrow a_{i_{l}}')_{i_{l}};\textbf{q}\rightarrow \textbf{q'})
\end{equation*}
This operator is defined as
\[u_{1}\otimes u_{2}\otimes\dots\otimes u_{p}\otimes (\textbf{q}\rightarrow \textbf{q'})\]
where $u_i = (a_{i_{j}} \rightarrow a'_{i_{j}})$ if $\exists j, i=i_{j}$, $u_{i}=1$ elsewhere.%, and $\textbf{q} \rightarrow \textbf{q}' = \text{Id}$ if $\textbf{q} = \textbf{q}'$.

Finally, we will denote by $\tau_{i,j}$ the operator in $\vn{G}$ induced by the transposition exchanging the integers $i$ and $j$.

\begin{figure*}
\begin{align*}
\text{bf}_{j,\textbf{q'}}^{c}&=\piout\otimes\tau_{0,j}\otimes((\{0o,1o,\star o\}\rightarrow)_{j};\textbf{q}\rightarrow \textbf{mov}^{\textbf{q,q'}}_{j})\\
\text{ff}_{j,\textbf{q'}}^{c}&=(\text{in}\rightarrow\text{out})\otimes\tau_{0,j}\otimes((\{0i,1i,\star i\}\rightarrow)_{j};\textbf{q}\rightarrow \textbf{mov}^{\textbf{q,q'}}_{j})\\
\text{fb}_{j,\textbf{q'}}^{c}&=\piin\otimes\tau_{0,j}\otimes((\{0i,1i,\star i\}\rightarrow)_{j};\textbf{q}\rightarrow \textbf{mov}^{\textbf{q,q'}}_{j})\\
\text{bb}_{j,\textbf{q'}}^{c}&=(\text{out}\rightarrow\text{in})\otimes\tau_{0,j}\otimes((\{0o,1o,\star o\}\rightarrow)_{j};\textbf{q}\rightarrow \textbf{mov}^{\textbf{q,q'}}_{j})\\
\text{rec}_{j,\textbf{q'}}^{c}&=\sum_{\bullet\in\{0,1,\star\}} \big( \pi_{\bullet i}\otimes\tau_{0,j}\otimes((\rightarrow \bullet i)_{j};\textbf{mov}^{\textbf{q,q'}}_{j}\rightarrow \textbf{q'}) \big)
\end{align*}
\caption{The operators in $\vn{M}_{6}(\complexN)\otimes\vn{G}\otimes\vn{P}$ encoding forward and backward move instructions.}\label{operateurs1}
\end{figure*}

\begin{figure*}
\begin{align*}
\text{rm}_{j}&=1\otimes\tau_{0,j}\otimes(\textbf{back}_{j}\rightarrow \textbf{mov-back}_{j})\\
\text{rr}^{0}_{j}&=\pi_{0\textnormal{o}}\otimes\tau_{0,j}\otimes((\rightarrow 0o)_{j};\textbf{mov-back}_{j}\rightarrow \textbf{back}_{j})\\
\text{rr}^{1}_{j}&=\pi_{1\textnormal{o}}\otimes\tau_{0,j}\otimes((\rightarrow 1o)_{j};\textbf{mov-back}_{j}\rightarrow \textbf{back}_{j})\\
\text{rc}_{j}&=\left\{\begin{array}{ll}\pi_{\star\textnormal{o}}\otimes\tau_{0,j}\otimes((\rightarrow a_{j})_{j};\textbf{mov-back}_{j}\rightarrow \textbf{back}_{j+1})&~~~(1\leqslant j<p)\\
\pi_{\star\textnormal{o}}\otimes\tau_{0,p}\otimes((\rightarrow a_{p})_{p};\textbf{mov-back}_{p}\rightarrow \textbf{q}_{0})&~~~(j=p)\end{array}\right.
\end{align*}
\caption{The operators in $\vn{M}_{6}(\complexN)\otimes\vn{G}\otimes\vn{P}$ encoding rejection.\newline
The operator $\text{rc}_{j}$ is parametric on an initial pseudo-configuration $s = (a_1, \hdots, a_p; \textbf{q}_0)$.
}\label{operateurs2}
\end{figure*}

Now, we can explain how the different transitions are encoded.
\Cref{operateurs1,operateurs2}
give the definitions of the operators used in the following description.
\paragraph{Move a pointer, read a value and change state}

We first explain how to encode the action \enquote{move forward the pointer $j$}. When we are in the pseudo-configuration $c=(a_{1},\dots,a_{p};\textbf{q})$, the machine, during this transition, moves the pointer $j$ one step to the right, reads the value $a'_{j}$ stored at $\#p_j$, updates the memory slot in consequence and changes the state from $\textbf{q}$ to $\textbf{q}'$. There are two cases: either the last movement of the pointer $j$ was a forward move, or it was a backward move. The case we are dealing with is obtained from the value stored in memory slot of the $j$-th pointer: if the value is $0i$, $1i$ or $\star i$, then the last move was a forward move, if the value is $0o$, $1o$ or $\star o$, the last move was a backward move. In the first case, the operator $\text{ff}_{j,\textbf{q'}}^{c}$ will be applied (notice the projection $\piin$ on the $j$-th memory slot), and in the second the $\text{bf}_{j,\textbf{q'}}^{c}$ operator will be applied. Both these operators somehow \emph{
activate} the $j$-th pointer by using the transposition $\tau_{0,j}$ and prepare for reading the representation of the integer. This representation will then give the value of the next (when moving forward) digit of the input. The $\text{rec}_{j,\textbf{q'}}^{c}$ operator is then applied in order to simultaneously update the value of the $j$-th memory slot and \emph{deactivate} the $j$-th pointer.

%We first explain how to encode the action ``move forward the pointer $j$ when we are in the pseudo-configuration $c=(a_{1},\dots,a_{p};\textbf{q})$, read the value $a'_{j}$ stored at $\#p_j$ and change the pseudo-configuration for $c'=(a_{1},\dots,a_{j-1},a'_{j},a_{j+1},\dots,a_{p};\textbf{q}')$''. There are two cases: either the last movement of the pointer $j$ was a forward move, either it was a backward move. The case we are dealing with is obtained from the value stored in memory slot of the $j$-th pointer: if the value is $0i$, $1i$ or $\star i$, then the last move was a forward move, if the value is $0o$, $1o$ or $\star o$, the last move was a backward move. In the first case, the operator $\text{ff}$ will be applied (notice the projection $\piin$ on the $j$-th memory slot), and in the second the $\text{bf}$ operator will be applied. Both these operators somehow \emph{activate} the $j$-th pointer by using the transposition $\tau_{0,j}$ and prepare for reading the representation of the integer. This representation will then give the value of the next (when moving forward) digit of the input. The $\text{rec}$ operator is then applied in order to simultaneously update the value of the $j$-th memory slot and \emph{disable} the $j$-th pointer.

%For $a \in\{f, b\}, g\in\{0,1, \star\}$, and\footnote{We consider that $\star o=\text{start}$ and $\star i=\text{end}$.} $d=gi$ (resp.\ $d = go$) if $a = f$ (resp.\ $a = b$), we define the operators:

The operator that will encode the moving forward instruction is then defined as $\text{forward}_{j,\textbf{q'}}^{c}=\text{bf}_{j,\textbf{q'}}^{c}+\text{ff}_{j,\textbf{q'}}^{c}+\text{rec}_{j,\textbf{q'}}^{c}$ which is an element of the algebra $\vn{M}_{6}(\complexN)\otimes\vn{G}\otimes\vn{P}$.

In case of a \enquote{move backwards pointer $j$} instruction, the operator encoding the transition is $\text{backward}_{j,\textbf{q'}}^{c}=\text{fb}_{j,\textbf{q'}}^{c}+\text{bb}_{j,\textbf{q'}}^{c}+\text{rec}_{j,\textbf{q'}}^{c}$, one again an element of the algebra $\vn{M}_{6}(\complexN)\otimes\vn{G}\otimes\vn{P}$.

In the following, we will forget about the subscripts and superscripts of these operators\footnote{Remark by the way that all those indices were redundant, for the pseudo-configuration \(c\) entails a choice of \(j\) and \(\textbf{q}'\).} and write $\text{ff}$, $\text{fb}$, $\text{bf}$, $\text{bb}$ and $\text{rec}$ in order to simplify notations.

\paragraph{Accept}
The case of acceptance is especially easy: we want to stop the computation\footnote{This is one of the reasons we did not considered \textbf{accept} to be a state: we want the computation to stop immediately, there is no need to provide \enquote{one last set of instructions}.}, so every transition $(a_{1},\dots,a_{n};\textbf{q})\rightarrow \textbf{accept}$ will be encoded by $0$.

\paragraph{Reject}
The transitions of the form $(a_{1},\dots,a_{n};\textbf{q})\rightarrow\textbf{reject}$ are encoded by the operator that represents the \enquote{transition} $(a_{1},\dots,a_{n};\textbf{q})\rightarrow(a_{1},\dots,a_{n};\textbf{back}_{1})$.

This defines the operator $\rightarrow^{\bullet}$ encoding the transition relation $\rightarrow$. But we need a bit more to interpret a machine. Indeed, as we already explained, the representation of an integer is accepted by an observation $\phi$ if the product of the two operators is nilpotent. Hence, a rejection should lead to the creation of a loop. We therefore add two new states -- $\textbf{back}_{j}$ and $\textbf{mov-back}_{j}$ -- for each $j=1,\dots,p$. We can then define $\text{reject}_{s}=\sum_{j=1}^{p} \text{rm}_{j} + \text{rr}^{0}_{j}+ \text{rr}^{1}_{j} + \text{rc}_{j}$. This operator actually encodes a reinitialization%
\footnote{And this is why the encoding of a NDPM is done provided an initial pseudo-configuration $s$. In practice, one can always take \(s\) to be the \enquote{faithful} pseudo-configuration, i.e.\ \(s = (\star, \dots, \star ; \textbf{q}_0)\). We choose to dwell on the parametricity of the encoding to highlight the particular mechanism of rejection, and its flexibility.}
 of the machine that starts when the computation reaches the state $\textbf{back}_{1}$: the pointers are moved back one after the other until they point on the symbol $\star$, and the memory slots and the state are initialized according to the given initial pseudo-configuration $s=(
a_{1},\dots, a_{p};\textbf{q}_{0})$. The reason for this complicated way of encoding rejection comes from the fact that the computation simulated by the operators is massively parallel: the operator $M_s^{\bullet}$ is a big sum that recollect every single evolution from a pseudo-configuration to any other \enquote{reachable} with $\to$, and it starts moreover in all pseudo-configurations at the same time.

Another complication due to this parallel computation and this particular way of defining rejection is that the operators simulate correctly only \emph{acyclic machines}, i.e.\ machines that never enter a computational loop whichever configuration -- including intermediary ones -- is chosen as a starting point for computation\footnote{For instance, let us consider a machine $M$ that contains a state $\textbf{loop}$ and, for a chosen pseudo-configuration $c$, the transition $(c,\textbf{loop})\rightarrow (\epsilon_1, \hdots, \epsilon_p,\textbf{loop})$. Then this machine is not acyclic, even if the state $\textbf{loop}$ cannot be accessed during the computation, i.e.\ even if $M$ do not contain any transition leading to the state $\textbf{loop}$.}. This is necessary because we want the entering of a loop to be the result of rejection. Fortunately, it can be shown that:

\begin{proposition}[{\cite[Lemma 25]{aubert2014ctemp}}]
\label{acyclicprop}
For all $M \in \text{NDPM}(p)$ there exists $M' \in \text{NDPM}(p')$ such that $M'$ is acyclic and $\mathcal{L}(M) = \mathcal{L}(M')$.
\end{proposition}

In the particular case of acyclic machines, one can show that the encoding is sound:
\begin{proposition}[{\cite[Lemma 29]{aubert2014ctemp}}]
Let $M \in \text{NDPM}(p)$ be acyclic, $c\in C_M$ and $M_s^{\bullet}=\rightarrow^{\bullet}+\textnormal{reject}_{s}$. For all $n \in \naturalN$ and every binary representation $N_{n}\in\vn{M}_{6}(\complexN)\otimes \vn{N}_{0}$ of the integer $n$:
\begin{equation*}
\text{$M_s(n)$ accepts}\Leftrightarrow\text{$M^{\bullet}_{s}(N_{n}\otimes 1)$ is nilpotent.}
\end{equation*}
\end{proposition}

These two results can then be combined with the results of the previous section to get the following proposition:
\begin{proposition}\label{incluspplus}
\[\cc{co-NL}\subseteq [P_{+}]\]
\end{proposition}

We will now study the encoding of deterministic pointer machines in more detail.

\subsection{The Encoding of DPMs}\label{encod-det}

We will show that the encoding of deterministic pointer machines satisfies a particular property: their $1$-norm is less or equal to $1$. We first define what is the $1$-norm on the algebras $\vn{M}_{k}(\vn{M}_{6}(\finhyp))$; this defines in particular the $1$-norm of observations since $\vn{M}_{6}(\complexN)\otimes\vn{G}\otimes\vn{P}$ is a subalgebra of $\vn{M}_{6}(\complexN)\otimes\finhyp\otimes\vn{M}_{k}(\complexN)$ (for $k=6^{p} \times \card{Q^{\uparrow}}$), which is isomorphic to $\vn{M}_{k}(\vn{M}_{6}(\finhyp))$.

\begin{definition}
We define the family of norms $\norm{\cdot}_{1}^{k}$ on $\vn{M}_{k}(\vn{M}_{6}(\finhyp))$ by:
\[\norm{(a_{i,j})_{1\leqslant i,j\leqslant k}}_{1}^{k}=\max_{1\leqslant j\leqslant k} \sum_{i=1}^{k}\norm{a_{i,j}}\]
where $\norm{a_{i,j}}$ is the usual norm (the C$^{\ast}$-algebra norm) on $\vn{M}_{6}(\finhyp)$.\\
We will simply write $\norm{\cdot}_{1}$ when the superscript is clear from the context.
\end{definition}

\begin{remark}
For each $k$, the map $\norm{\cdot}_{1}^{k}:\vn{M}_{k}(\vn{M}_{6}(\finhyp))\rightarrow\realposN$ just defined is a norm and, as such, does not depend on the chosen basis. To understand this, notice that this norm can be defined -- considering the isomorphism between $\vn{M}_{k}(\vn{M}_{6}(\finhyp))$ and $\vn{M}_{k}(\complexN)\otimes\vn{M}_{6}(\finhyp)$ -- as the tensor product of the $1$-norm on $\vn{M}_{k}(\complexN)$ and the usual norm on $\vn{M}_{6}(\finhyp)$. The $1$-norm on $\vn{M}_{k}(\complexN)$ is formally defined from the usual $1$-norm on vectors as $\norm{A}_{1}=\sup \frac{\norm{Ax}_{1}}{\norm{x}_{1}}$. It can then be shown equal, for any (orthonormal) basis in which one may write $A$ as a matrix $(a_{i,j})_{1\leqslant i,j\leqslant k}$, to the expression $\max_{1\leqslant j\leqslant k} \sum_{i=1}^{k} \abs{a_{i,j}}$.
%The $1$-norm on $\vn{M}_{k}(\complexN)$ is formally defined from the usual $1$-norm on vectors as $\norm{A}_{1}=\sup \frac{\norm{Ax}_{1}}{\norm{x}_{1}}$.
%Let us consider an (orthonormal) basis in which one may write $A$ as a matrix $(a_{i,j})_{1\leqslant i,j\leqslant k}$.
%Then one can prove that the $1$-norm on $\vn{M}_{k}(\complexN)$ can be shown equal to the expression $\max_{1\leqslant j\leqslant k} \sum_{i=1}^{k} \abs{a_{i,j}}$, where \(\abs{\cdot}\) is ...
\end{remark}

A corollary of the following proposition will be used extensively in the proofs.

\begin{proposition}\label{propprincnorm}
Let $A,B$ be operators such that $AB^{\ast}=B^{\ast}A=0$. Then:
\[\norm{A+B}=\max\{\norm{A},\norm{B}\}\]
\end{proposition}

\begin{proof}
Actually, this proposition appears as Theorem 1.7 (b) in an article by P.J. Maher~\cite{maher1990}. This last theorem is stated differently (with conditions on ranges) in the given reference. To obtain the result as we stated it, simply notice that $AB^{\ast}=0$ implies $\ran(B^{\ast})\subseteq \ker(A)=(\ran(A^{\ast}))^{\bot}$ so $\ran(B^{\ast})\perp\ran(A^{\ast})$. Similarly, $B^{\ast}A=0$ implies that $\ran(A)\subseteq\ker(B^{\ast})=(\ran(B))^{\bot}$ so $\ran(A)\perp\ran(B)$. Thus, if $AB^{\ast}=B^{\ast}A=0$, we have $\ran(A)\perp\ran(B)$ and $\ran(A^{\ast})\perp\ran(B^{\ast})$ and one can then apply the theorem of Maher.
\end{proof}

\begin{corollary}\label{cornorm}
The operators $\text{ff}+\text{bf}$, $\text{rec}$, $\text{bb}+\text{fb}$, and $\text{rr}^{0}_{j}+\text{rr}^{1}_{j}+\text{rc}_{j}$ ($j$ fixed) are of norm $1$.
\end{corollary}

\begin{proof}
It is clear that $\text{bf}\times\text{ff}^{\ast}=0$ since $\piout\piin=0$. Similarly, $\text{ff}^{\ast}\times\text{bf}=0$. Thus, using the preceding proposition, we have $\norm{\text{ff}+\text{bf}}=\max\{\norm{\text{bf}},\norm{\text{ff}}\}=1$. 

A similar argument shows that $\norm{\text{fb}+\text{bb}}=1$.

Clearly, as a consequence of $\pi_{ai}\pi_{bi}=0$ for $a\neq b$, we deduce $\norm{\text{rec}}=1$ by \cref{propprincnorm}.

We are now proving the result for $\text{rr}^{0}_{j}+\text{rr}^{1}_{j}+\text{rc}_{j}$, so we fix $1\leqslant j \leqslant p$. First, notice that $\text{rc}_{j}\times(\text{rr}^{0}_{j})^{\ast}=0$ as a consequence of $\pi_{\textbf{mov-back}_{j}}\times\pi_{\textbf{back}_{j}}=0$. Conversely, $(\text{rr}^{0}_{j})^{\ast}\times\text{rc}_{j}=0$ as a consequence of\footnote{When $j=p$, we consider that $\textbf{back}_{j+1}=\textbf{q}_{0}$ to ease the argument.} $\pi_{\textbf{back}_{j+1}}\times\pi_{\textbf{mov-back}_{j}}=0$. A similar argument shows that $\text{rc}_{j}\times(\text{rr}^{1}_{j})^{\ast}=0$ and $(\text{rr}^{1}_{j})^{\ast}\times\text{rc}_{j}=0$. Moreover, we know that $(\text{rr}^{0}_{j})^{\ast}\times\text{rr}^{1}_{j}=0$ and $\text{rr}^{1}_{j}\times (\text{rr}^{0}_{j})^{\ast}=0$ by just looking at the first term of the tensor in their definition. By an iterated use of \cref{propprincnorm}, we get that $\norm{\text{rr}^{0}_{j}+\text{rr}^{1}_{j}+\text{rc}_{j}}=\max\{\norm{\text{rr}_{j}^{0}},\norm{\text{rr}_{j}^{1}},\norm{\text{rc}_{j}}\}=1$.
\end{proof}

We are now in possession of all the material needed to characterize the operators coming from the encoding of a deterministic pointer machine.

\begin{proposition}\label{prop1norm}
Let $M\in \text{DPM}(p)$ be an acyclic machine, and $M_s^{\bullet}=\rightarrow^{\bullet}+\text{reject}_{s}$ its encoding as an operator, $\norm{M^{\bullet}_{s}}_{1}\leqslant 1$.
\end{proposition}

\begin{proof}
Since we are working with deterministic machines, the transition relation is functional: for each $\rho\in C_{M}$ there is at most one $t$, say $t_{\rho}$, such that $\rho\rightarrow t_{\rho}$. Thus:
\[M_{s}^{\bullet}=\sum_{\rho\in C_{M}} \phi_{\rho,t_{\rho}}+\text{reject}_{s}\]
Since $M_{s}^{\bullet}$ is an element of $\vn{M}_{k}(\vn{M}_{6}(\finhyp))$, we will now compute $\norm{M^{\bullet}_{s}}_{1}^{k}$. We first show that this matrix has at most one non-null coefficient in each column. Then we will use the fact that these coefficients are of norm at most $1$.

%\note{We will write the set of \emph{extended pseudo-configurations} as the set $P=\{1,\dots,6\}^{p}\times Q^{\uparrow}$, which is a basis of the algebra of pseudo-states, and we write $M^{\bullet}_{s}=(a_{\rho,\rho'})_{\rho,\rho'\in P}$. $\to$ 
To prove this we introduce the set $P=\{0i, 0o, 1i, 1o, \star i, \star o\}^{p}\times Q^{\uparrow}$ of \emph{extended pseudo-configurations}. This set is a basis of the algebra of pseudo-states, and we write $M^{\bullet}_{s}=(a_{\rho,\rho'})_{\rho,\rho'\in P}$. This way, the $1$-norm of $M^{\bullet}_{s}$ is
\[\norm{M^{\bullet}_{s}}_{1}=\max_{\rho'\in P}\{\sum_{\rho\in P} \norm{a_{\rho,\rho'}}\}\] 
Let $\rho=(a_{1},\dots,a_{p}; q)$ be a (non-extended) pseudo-configuration. If $\rho$ is a pseudo-configuration of the machine (i.e.\ $q$ is an element of $Q$), then there is at most one pseudo-configuration $t_{\rho}$ such that $\rho\rightarrow t_{\rho}$. This \emph{atomic transition} can be:
\begin{itemize}
\item Accept: in this case the operator $\phi_{\rho,t_{\rho}}$ is equal to $0$ and the column is empty.
\item Reject: in this case the column corresponding to $\rho$ contains only the operator $(\textbf{q}\rightarrow\textbf{back}_{1})$ that encodes the transition $\rho\rightarrow (a_{1},\dots,a_{p};\textbf{back}_{1})$, and the norm of this operator is equal to $1$.
\item Move forward a pointer, read a value and change state to $\textbf{q'}$: then the only extended pseudo-configurations introduced by the encoding is $\tilde{\rho}=(a_{1},\dots,a_{n}; \textbf{mov}_{j}^{\textbf{q,q'}})$. The column corresponding to $\rho$ contains the operator $\text{ff}+\text{bf}$, which is of norm $1$ by \cref{cornorm}. The column corresponding to $\tilde{\rho}$ contains only the operator $\text{rec}$ whose norm is equal to $1$ by \cref{cornorm}.
\item Move backwards: this case is similar to the previous one.
\end{itemize}
Now let us take a look at the operator $\text{reject}_{s}$. The extended pseudo\hyp{}configurations introduced for the encoding of the rejection are, for $j=1,\dots,p$, $\bar{\rho}^{m}_{j}=(a_{1},\dots,a_{n}; \textbf{mov-back}_{j})$ and $\bar{\rho}^{b}_{j}=(a_{1},\dots,a_{n}; \textbf{back}_{j})$. The column corresponding to $\rho$ contains only the operator encoding the transition from $\rho$ to $\bar{\rho}_{1}^{b}$, which is a norm $1$ operator. Let us now fix $1\leqslant j \leqslant p$. The column corresponding to the extended pseudo-state $\bar{\rho}_{j}^{b}$ contains only the operator $\text{rm}_{j}$, which is of norm $1$. The column corresponding to $\bar{\rho}_{j}^{m}$ contains the operator $\text{rr}_{j}+\text{rc}_{j}$, which is of norm $1$ by \cref{cornorm}.

We just showed that each column of the matrix contains at most one operator different from $0$, and that this operator is always of norm $1$. Hence, for any fixed extended pseudo-configuration $\rho$:
\begin{equation*}
\sum_{\rho\in P} a_{\rho,\rho'}\leqslant 1
\end{equation*}
As a consequence of the definition of the $1$-norm, we have:
\[\norm{M^{\bullet}}_{1}=\max_{\rho'\in P}\{\sum_{\rho\in P} a_{\rho,\rho'}\}\leqslant 1\]
\end{proof}

The last proposition motivates the following definition:

\begin{definition}
\[P_{+,1}=\{\phi \mid \phi\in P_{+}\textnormal{ and }\norm{\phi}_{1}\leqslant 1\}\]
\end{definition}

Now, by looking carefully at the proof of \cref{acyclicprop}, one can show that a similar result holds in the particular case of deterministic pointer machines. Indeed, if $M$ is deterministic, the machine $M'$ constructed from $M$ in the proof is also deterministic:

\begin{proposition}
For all $M \in\text{DPM}(p)$, there exists $M' \in\text{DPM}(p)$ such that $M'$ is acyclic and $\mathcal{L}(M) = \mathcal{L}(M')$.
\end{proposition}

As a consequence, one gets the following result:
\begin{proposition}
\[\cc{L}\subseteq [P_{+,1}]\]
\end{proposition}

The question is then: is this inclusion strict or did we find a characterization of the class $\cc{L}$? To answer this question, we need to look at the proof of the inclusion $[P_{\geqslant 0}]\subseteq \cc{co-NL}$ and see what happens when we restrict to the operators in $P_{+,1}$.

\section{Operators and Logarithmic Space}

In this section, we will show that $[P_{+,1}]\subseteq \cc{L}$, concluding the proof that $[P_{+,1}]=\cc{L}$. To prove this inclusion, we need a technical lemma, which we proved in our earlier paper.

\begin{lemma}[{\cite[Lemma~31]{aubert2014ctemp}}]\label{technicallemma}
Let $N_{n}$ be a binary representation of an integer $n$ in $\vn{M}_{6}(\complexN)\otimes\vn{N}_{0}$ and $\Phi\in\vn{M}_{6}(\complexN)\otimes\vn{S}\otimes\vn{M}_{k}(\complexN)$ be an observation in $P_{\geqslant 0}$. There exist an integer $f$, an injective morphism $\psi:\vn{M}_{f}(\complexN)\rightarrow\finhyp$ and two matrices $M\in\vn{M}_{6}(\complexN)\otimes\vn{M}_{f}(\complexN)$ and $\bar{\Phi}\in\vn{M}_{6}(\complexN)\otimes\vn{M}_{f}(\complexN)\otimes\vn{M}_{k}(\complexN)$ such that $\text{Id}\otimes\psi(M)=N_{n}$ and $\text{Id}\otimes\psi\otimes\text{Id}_{\vn{E}}(\bar{\Phi})=\Phi$.
\end{lemma}

This lemma is of the utmost importance. Even though the hyperfinite factor $\finhyp$ is necessary to have a uniform representation of integers, it is an algebra of operators acting on an infinite-dimensional Hilbert space. It is therefore not obvious that one can check the nilpotency of such an operator with finite resources, and it is of course not possible in general. 

However, the technical lemma above has as a direct consequence that checking the nilpotency of a product $\Phi (N_{n}\otimes 1)$ where $\Phi$ is an observation in $P_{\geqslant 0}$ is equivalent to checking the nilpotency of a product of matrices.

\begin{corollary}
Let $\Phi\in\vn{M}_{6}(\complexN)\otimes\vn{G}\otimes\vn{M}_{k}(\complexN)$ be an observation and $N_{n}\in\vn{M}_{6}(\complexN)\otimes\vn{N}_{0}$ a representation of the integer $n$. There exists \emph{matrices} (i.e.\ operators acting on a finite-dimensional Hilbert space) $\bar{\Phi}$ and $M$ such that:
\[\Phi (N_{n}\otimes 1_{\vn{M}_{k}(\complexN)})\text{ is nilpotent if and only if $\bar{\Phi}(M\otimes 1_{\vn{M}_{k}(\complexN)})$ is nilpotent}\]
\end{corollary}

The matrices $\bar{\Phi}$ and $M\otimes1_{\vn{M}_{k}(\complexN)}$ are acting on a Hilbert space of dimension $6\times (\log_{2}(n)+1)^{p}p! \times k$, i.e. the integer $f$ in \Cref{technicallemma} is equal to $(\log_{2}(n)+1)^{p}p!$. Moreover the degree of nilpotency of the two products is equal. We can then show that there exists a Turing machine that decides the nilpotency of the product $\Phi N_{n}$ using only logarithmic space (i.e.\ in $\log_{2}(n)$).

\begin{proposition}[{\cite[Proposition 32]{aubert2014ctemp}}]
\label{boincluconl}
If $\Phi\in P_{\geqslant 0}$ and $N_{n}$ is a representation of $n\in\naturalN$, there is a \cc{co-NL} machine that checks if $\bar{\Phi}(M\otimes 1_{\vn{M}_{k}(\complexN)})$ is nilpotent.
\end{proposition}

This proposition, together with \cref{incluspplus} and the fact that $[P_{+}]\subseteq [P_{\geqslant 0}]$ gives us a proof of the following equality:
\begin{theorem}[{\cite[Theorem 33]{aubert2014ctemp}}]\label{thmprincconl}
\[\cc{co-NL}=[P_{+}]=[P_{\geqslant 0}]\]
\end{theorem}

We recall how the proof of \cref{boincluconl} works. The algebra we are working with is:
\[\vn{F}=\vn{M}_{6}(\complexN)\otimes((\underbrace{\vn{M}_{\log_{2}(n)+1}(\complexN)\otimes\dots\otimes\vn{M}_{\log_{2}(n)+1}(\complexN)}_{p\text{ copies}})\rtimes\mathfrak{S}_{p})\otimes\vn{M}_{k}(\complexN)\]
We let in the following $K = 6k(\log_{2}(n)+1)^{p}p!$ be the dimension of the Hilbert space the algebra $\vn{F}$ is acting on.

We chose the basis of this algebra defined as the set of elements of the form:
\[(\pi,a_{0},a_{1},\dots,a_{p};\sigma;e)\]
where $\pi$ is an element of the basis $(0o,0i,1o,1i,s,e)$ of $\vn{M}_{6}(\complexN)$, $a_{i}$ ($i\in\{1,\dots,p\}$) are the elements of the basis chosen to represent the integer $n$, $\sigma \in \mathfrak{S}_{p}$ and $e$ is an element of a basis of $\vn{M}_{k}(\complexN)$. When we apply $M\otimes 1_{\vn{M}_{k}(\complexN)}$ representing the integer to an element of this basis, we obtain one and only one vector of the basis $(\pi,a_{0},a_{1},\dots,a_{p};\sigma;e)$. When we apply to this element the matrix $\bar{\Phi}$ we obtain a (positive) linear combination of $l\in\naturalN$ elements of the basis:
\begin{equation}
\label{eqphibar}
\bar{\Phi}(\pi,a_{0},a_{1},\dots,a_{p};\sigma;e)=\sum_{i=0}^{l}\alpha_{i}(\rho,a_{\tau_{i}(0)},\dots,a_{\tau_{i}(p)};\tau_{i}\sigma;e_{i})
\end{equation}

So we obtain the following picture:
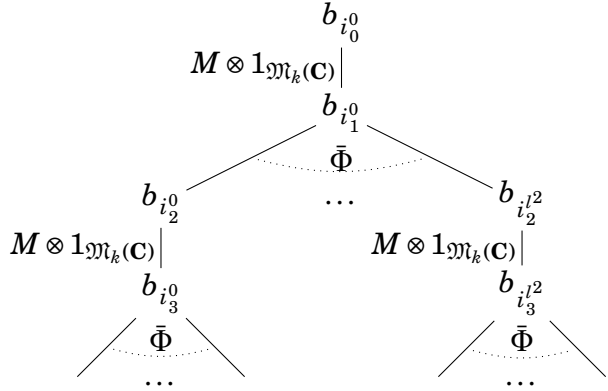
\begin{center}
\begin{tikzpicture}[x=1.2cm,y=0.6cm]
	\node (R) at (0,0) {$b_{i^{0}_{0}}$};
	\node (N1) at (0,-2) {$b_{i^{0}_{1}}$};
	\node (P1) at (-2,-4) {$b_{i^{0}_{2}}$};
	\node (P2) at (2,-4) {$b_{i^{l^{2}}_{2}}$};
	\node (Pmid) at (0,-4) {$\dots$};
		\node (V) at (0,-3) {$\bar{\Phi}$}; 
	\node (N11) at (-2,-6) {$b_{i^{0}_{3}}$};
	\node (N22) at (2,-6) {$b_{i^{l^{2}}_{3}}$};
	\node (P11) at (-3,-8) {};
	\node (Pmid1) at (-2,-8) {$\dots$};
		\node (U) at (-2,-7) {$\bar{\Phi}$}; 
	\node (P12) at (-1,-8) {};
	\node (P21) at (1,-8) {};
	\node (Pmid2) at (2,-8) {$\dots$};
		\node (T) at (2,-7) {$\bar{\Phi}$}; 
	\node (P22) at (3,-8) {};
	
	\draw[-] (R) -- (N1) node [midway,left] {$M\otimes 1_{\vn{M}_{k}(\complexN)}$};
	\draw[-] (N1) -- (P1) node [midway] (A) {};
	\draw[-] (N1) -- (P2) node [midway] (B) {};
	\draw[-] (P1) -- (N11) node [midway,left] {$M\otimes 1_{\vn{M}_{k}(\complexN)}$};
	\draw[-] (P2) -- (N22) node [midway,left] {$M\otimes 1_{\vn{M}_{k}(\complexN)}$};
	\draw[-] (N11) -- (P11) node [midway] (A1) {};
	\draw[-] (N11) -- (P12) node [midway] (B1) {};
	\draw[-] (N22) -- (P21) node [midway] (A2) {};
	\draw[-] (N22) -- (P22) node [midway] (B2) {};
	
	\draw[dotted] (A) to [bend right=15] (B) {};
	\draw[dotted] (A1) to [bend right=15] (B1) {};
	\draw[dotted] (A2) to [bend right=15] (B2) {};
\end{tikzpicture}
\end{center}

To decide if the product $\bar{\Phi}(M\otimes 1_{\vn{M}_{k}(\complexN)})$ is nilpotent, it is then sufficient to check, for each possible value of $b_{i_{0}^{0}}$, that the branches of this tree are finite. Since they are either infinite or of length at most $K$, this can be checked by a non-deterministic (to deal with the branchings that appears in the figure above) Turing machine. This machine only uses logarithmic space since it needs only to store at each step the value of $b_{i_{0}^{0}}$, the current basis element $b_{i_{2j}^{h}}$, and the number of iterations $j$ of $\bar{\Phi}(M\otimes 1_{\vn{M}_{k}(\complexN)})$ that were computed to get from $b_{i_{0}^{0}}$ to $b_{i_{2j}^{h}}$.

Now, let us look at the case when $\Phi\in P_{+,1}$. In this particular case, \cref{eqphibar} becomes:
\begin{equation}
\label{eqphibardet}
\bar{\Phi}(\pi,a_{0},a_{1},\dots,a_{p};\sigma;e)=(\rho,a_{\tau_{b_{0}}(0)},\dots,a_{\tau_{b_{0}}(p)};\tau_{b_{0}}\sigma;e_{b_{0}})
\end{equation}
Let us write $\bar{\Phi}$ as a $\card{Q}\times\card{Q}$ matrix $(\bar{\Phi}_{q,q'})_{q,q'\in Q}$ (here $Q$ denotes a basis of $\vn{M}_{k}(\complexN)$, which is a matrix algebra from the definition of observations). We can deduce from $\norm{\bar{\Phi}}_{1}\leqslant 1$ that for all $q\in Q$, $\sum_{q'\in Q} \norm{\bar{\Phi}_{q,q'}}\leqslant 1$.

As $\Phi$ is an element of $P_{+}$, the matrix $\bar{\Phi}$ satisfies the following equation:
\begin{equation}
\label{eqphibarplus}
\bar{\Phi}(\pi,a_{0},a_{1},\dots,a_{p};\sigma;e)=\sum_{i=0}^{l}(\rho,a_{\tau_{i}(0)},\dots,a_{\tau_{i}(p)};\tau_{i}\sigma;e_{i})
\end{equation}
We deduce that $\norm{\bar{\Phi}_{e,e_{i}}}\geqslant 1$ ($i=1,\dots,l$), and therefore $\sum_{e'\in Q} \norm{\bar{\Phi}_{e,e'}}\geqslant l$. 
Since $\Phi\in P_{+,1}$, we know that $\norm{\bar{\Phi}_{q,q'}}\leqslant 1$ from which we deduce that $l\leqslant 1$.

That is, the previous picture becomes:
\begin{center}
\begin{tikzpicture}[x=1.2cm,y=0.6cm]
	\node (R) at (0,0) {$b_{i^{0}_{0}}$};
	\node (N1) at (0,-2) {$b_{i^{0}_{1}}$};
	\node (Pmid) at (0,-4) {$b_{i^{0}_{2}}$};
	\node (Nmid) at (0,-6) {$b_{i^{0}_{3}}$};
	\node (PPmid) at (0,-8) {};
	\node (Pmid1) at (0,-8.5) {$\vdots$};
		
	\draw[-] (R) -- (N1) node [midway,left] {$M\otimes 1_{\vn{M}_{k}(\complexN)}$};
	\draw[-] (Pmid) -- (Nmid) node [midway,left] {$M\otimes 1_{\vn{M}_{k}(\complexN)}$};
	\draw[-] (N1) -- (Pmid) node [midway,left] {$\bar{\Phi}$};
	\draw[-] (Nmid) -- (PPmid) node [midway,left] {$\bar{\Phi}$};
	
\end{tikzpicture}
\end{center}

Notice that the nilpotency degree of $\bar{\Phi}(M\otimes 1_{\vn{M}_{k}(\complexN)})$ is again at most $K$. One can thus easily define a deterministic Turing machine that takes the basis elements one after the other and compute the sequence $b_{i^{0}_{0}}$, $b_{i^{0}_{1}}$, $\dots$: if the sequence stops before the $K$-th step, the machine starts with the next basis element as $b_{i^{0}_{0}}$, and if the sequence did not stop at step $K$ it means the matrix was not nilpotent. This Turing machine needs no more than logarithmic space since it stores only the current starting basis element (i.e.\ $b_{i^{0}_{0}}$), the last computed term of the sequence $b_{i^{0}_{2j}}$ and the number of iterations computed so far $j$. We thus obtain the following proposition.

\begin{proposition}
If $\Phi\in P_{+,1}$ and $N_{n}$ is a representation of $n\in\naturalN$, there is a $\cc{L}$ machine that checks if $\bar{\Phi}M$ is nilpotent.
\end{proposition}

Putting together the results of this section and the last, we can finally state the main theorem of this paper:
\begin{theorem}
\[\cc{L}=[P_{+,1}]\]
\end{theorem}

Together with \cref{thmprincconl}, we thus obtained two sets of observations $P_{+}$ and $P_{+,1}$, satisfying $P_{+,1}\subsetneq P_{+}$ and such that $\cc{L}=[P_{+,1}]\subseteq[P_{+}]=\cc{co-NL}$.

\section{Conclusion}

This work both completes and extends the results obtained in our earlier paper~\cite{aubert2014ctemp} where we showed that the approach recently proposed by Girard~\cite{girard2012} for studying complexity classes succeeds in characterizing the complexity class \cc{co-NL}. On the one hand, we showed that the non-deterministic pointer machines, which were designed to mimic the computational behavior of operators, are really close to a well-known abstract machine, two-way multi-head finite automata. This result gives a much better insight on the pointer machines, and we hope this will help us in extending the model of pointer machines to study others complexity classes. It moreover strengthens the proof of the fact that the set of operators $P_{+}$ characterizes the class $\cc{co-NL}$. On the other hand, we extended the result by finding a subset of $P_{+}$, defined through a condition on the norm, that corresponds to the encoding of deterministic pointer machines. We showed first that deterministic pointer 
machines were equivalent to the notion of deterministic two-way multi-head finite automata, hence that $\cc{L}$ is a subset of the language associated to $P_{+,1}$. We then showed that the converse inclusion holds, showing that the set of operators $P_{+,1}$ is indeed a characterization of the complexity class $\cc{L}$.

The notion of acceptance used in this work (nilpotency) is closely related~\cite{seiller2012b} to the notion of interaction in Girard's geometry of interaction in the hyperfinite factor (GoI5). This should leads to the possibility of defining types in GoI5 corresponding to complexity classes. For this reason, and the fact that the GoI5 construction allows more classical implicit computational complexity approaches such as defining constrained exponential connectives, the GoI5 framework seems a perfect candidate for a general mathematical framework for the study of complexity. The combined tools offered by geometry of interaction and the numerous tools and invariants of the theory of operators that could be used in this setting offer new perspectives for the obtention of separation results.

We also believe that the approach explored in this paper can be used to obtain characterizations of other complexity classes, and in particular the classes \cc{P} and \cc{co-NP}. These characterizations may be obtained through the use of a more complex group action in the crossed product construction, or by defining a suitable superset of $P_{+}$. An approach for obtaining such a result would be to generalize the notion of pointer machines to get a characterization of $\cc{P}$ or $\textbf{co-NP}$. Since the pointer machines are closely related to the operators and the way the latter interact with the representation of integers, such a result would be a great step towards a characterization of these classes.

%\section*{Acknowledgments}
%Apart from the leaders of the ANR projects listed at the first page, the authors would like to thank the referee for providing remarks that considerably improved this paper.
%The authors also wish to warmly thank the editors of this special issue for their tremendous work.

%\section*{References}
%\addcontentsline{toc}{section}{References}
\bibliographystyle{elsarticle-num}
%\bibliography{../biblio}
\bibliography{nouvelle-bib}

\appendix
\def\appendixname{Appendix }

\section{Examples}\label{example}
\def\appendixname{}
We develop below some examples that should enlighten the encoding of integers, the notion of pointer machine, and the encoding of the latter as operators.
We begin by recalling the encoding of the input adopted, with two examples.
Then, we briefly sketch a pointer machine that accepts palindromes and encode it as an observation.
This allows us to make general remarks on the \enquote{tricks} one can adopt to lighten the encoding, which is in all generality unnecessarily heavy. Although this makes this example somehow strange to illustrate the paper as it is not encoded using the method employed for the proofs, we believe it gives a number of insights on how the observations interact with integers and how the computation actually takes place.
%We conclude by computing \enquote{by hand} the iteration of the representation of the two inputs given with the obtained observation.

%We will now define a specific pointer machine, explain how this pointer machine can be encoded as an observation, and we will finally compute the iterations of the encoding with the two integers just given as examples.

\subsection{Integers}

As explained in details in our earlier paper~\cite{aubert2014ctemp}, integers are represented as matrices which are then embedded into the hyperfinite factor $\finhyp$. The matrices are themselves adjacency matrices of a graph representing the links between two adjacent symbols in the binary representation of the integer. Let us work out two examples. 

First, the integer $\star 010$ will be represented by the adjacency matrix of the following graph. Notice the integers on top of the circled symbols which represent a sort of \enquote{state} -- or sorts of \enquote{locations} for symbols -- , with a unique such state for each symbol in the list (counting the symbol $\star$).
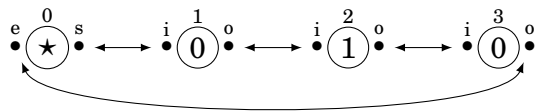
\begin{center}
\begin{tikzpicture}[inner sep=2pt, outer sep=1pt]
	\node[shape=circle,draw] (S) at (0,0) {$\star$};
		\node (SE) at (S.east) {};
		\node (SW) at (S.west) {};
		\node (SN) at (S.north) {};
		\node (So) at (SE.east) {$\bullet$};
		\node (SoL) at ($(So.north)+ (0,.05)$) {\scriptsize{s}};
		\node (Si) at (SW.west) {$\bullet$};
		\node (SiL) at ($(Si.north)+ (0,.05)$) {\scriptsize{e}};
		\node (St) at (SN.north) {\scriptsize{$0$}};
	\node[shape=circle,draw] (A) at (2,0) {$0$};
		\node (AE) at (A.east) {};
		\node (AW) at (A.west) {};
		\node (AN) at (A.north) {};
		\node (Ao) at (AE.east) {$\bullet$};
		\node (AoL) at ($(Ao.north)+ (0,.05)$) {\scriptsize{o}};
		\node (Ai) at (AW.west) {$\bullet$};
		\node (AiL) at ($(Ai.north)+ (0,.05)$) {\scriptsize{i}};
		\node (At) at (AN.north) {\scriptsize{$1$}};
	\node[shape=circle,draw] (B) at (4,0) {$1$};
		\node (BE) at (B.east) {};
		\node (BW) at (B.west) {};
		\node (BN) at (B.north) {};
		\node (Bo) at (BE.east) {$\bullet$};
		\node (BoL) at ($(Bo.north)+ (0,.05)$) {\scriptsize{o}};
		\node (Bi) at (BW.west) {$\bullet$};
		\node (BiL) at ($(Bi.north)+ (0,.05)$) {\scriptsize{i}};
		\node (Bt) at (BN.north) {\scriptsize{$2$}};
	\node[shape=circle,draw] (C) at (6,0) {$0$};
		\node (CE) at (C.east) {};
		\node (CW) at (C.west) {};
		\node (CN) at (C.north) {};
		\node (Co) at (CE.east) {$\bullet$};
		\node (CoL) at ($(Co.north)+ (0,.05)$) {\scriptsize{o}};
		\node (Ci) at (CW.west) {$\bullet$};
		\node (CiL) at ($(Ci.north)+ (0,.05)$) {\scriptsize{i}};
		\node (Ct) at (CN.north) {\scriptsize{$3$}};
	\draw[<->] (So) to (Ai) {};
	\draw[<->] (Ao) to (Bi) {};
	\draw[<->] (Bo) to (Ci) {};
	\draw[<->] (Co) .. controls (6,-1) and (0,-1) .. (Si) {};
\end{tikzpicture}
\end{center}
This adjacency matrix is a $6\times 6$ matrix with $4\times 4$ matrices as coefficients, where the number $4$ stands for the length of the list (these are the \enquote{states} mentioned above) and the number $6$ stand for the $6$ different types of vertices: $0i$, $0o$, $1i$, $1o$, $e$ (i.e.\ \(\star i\)) and $s$ (i.e.\ \(\star o\)). Here is the corresponding matrix:
\[
M_{\star 010}=
\begin{blockarray}{ccccccc}
\BAmulticolumn{2}{c}{{\overbrace{~~~~~~~~~~~~~~}^0}} & \BAmulticolumn{2}{c}{{\overbrace{~~~~~~~~~~~~~~}^1}}& \BAmulticolumn{2}{c}{{\overbrace{~~~~~~~~~~~~~~}^{\ast}}} \\
\begin{block}{(cccccc)c}
	0 & 0 & 0 & l_{10} & s_{0} & 0 & \multirow{2}{*}{$\left.\begin{tabular}{c} \hspace{-0.5cm} \vspace{0.4cm} \end{tabular}\right\}\scriptstyle{0}$}\\
	0 & 0 & l_{01}^{\ast} & 0 & 0 & e_{0}^{\ast}& \\
	0 & l_{01} & 0 & 0 & 0 & 0 & \multirow{2}{*}{$\left.\begin{tabular}{c} \hspace{-0.5cm} \vspace{0.4cm} \end{tabular}\right\} \scriptstyle{1}$}\\
	l_{10}^{\ast} & 0 & 0 & 0 & 0 & 0 & \\
	s_{0}^{\ast} & 0 & 0 & 0 & 0 & 0 & \multirow{2}{*}{$\left.\begin{tabular}{c} \hspace{-0.5cm} \vspace{0.4cm} \end{tabular}\right\} \scriptstyle{*}$}\\
	0 & e_{0} & 0 & 0 & 0 & 0 & \\
\end{block}
\end{blockarray}
\]
where $0$ stands for the zero $4\times 4$ matrix, the adjoint $(\cdot)^{\ast}$ corresponds to the conjugate-transpose, and the matrices $l_{10}$, $l_{01}$, $s_{0}$ and $e_{0}$ are defined as follows:
\[
l_{01} = \left(\begin{array}{cccc}
	0 & 0 & 0 & 0\\
	0 & 0 & 0 & 0\\
	0 & 1 & 0 & 0\\
	0 & 0 & 0 & 0
\end{array}\right)
l_{10} = \left(\begin{array}{cccc}
	0 & 0 & 0 & 0\\
	0 & 0 & 0 & 0\\
	0 & 0 & 0 & 0\\
	0 & 0 & 1 & 0
\end{array}\right)
\]
\[
s_{0} = \left(\begin{array}{cccc}
	0 & 0 & 0 & 0\\
	1 & 0 & 0 & 0\\
	0 & 0 & 0 & 0\\
	0 & 0 & 0 & 0
\end{array}\right)
e_{0} = \left(\begin{array}{cccc}
	0 & 0 & 0 & 1\\
	0 & 0 & 0 & 0\\
	0 & 0 & 0 & 0\\
	0 & 0 & 0 & 0
\end{array}\right)
\]
The matrix \(l_{01}\) express the edges of the above graph that are going from left to right, whose source is a $0$ and whose target is a $1$. This matrix is therefore located -- in the $6\times 6$ matrix -- on the column corresponding to \enquote{$0o$} (for \enquote{$0$ out}) -- because its source is a vertex \enquote{$0o$} -- and on the row \enquote{$1i$} -- because its target is a \enquote{$1i$} vertex. This matrix then contains a single non-zero element because the list contains exactly one such edge going left-to-right from a $0$ to a $1$, and shows how this edge connects the locations; in this case, the only such edge goes from the location named $1$ to the location named $2$, which explains why the non-zero element is located in the second column (locations are numbered starting from $0$) and the third row.
%This corresponds to the non-null value at the intersection of the second column and third row (remember that we have to keep a location numbered \(0\) for \(\star\)).
%It recollects all the location where it is possible to go from right to left from a \(0\) to a \(1\).

The definition and placement of the matrix $l_{10}$ is explained in the same fashion, while the matrix $s_{0}$ (resp.\ $e_{0}$) represent those edges going from left to right whose source is a \enquote{$s$} vertex (resp.\ a \enquote{$0$} vertex) and target is a \enquote{$0$} vertex (resp.\ a \enquote{$e$} vertex). Similar matrices named \(l_{00}\) and \(l_{11}\) could appear in the encoding of an integer if its binary writing contained two symbols $0$ or two symbols $1$ following each other; in our example, however, these matrices are empty.

Lastly, the adjoint matrices such as $l_{01}^{\ast}$ actually correspond to the right-to-left edge in the above graph.
%The matrices \(s_0\) and \(e_0\) simply expresses that the value connected to them are \(0\) and \(0\).

As a second example, we consider the integer $\star 1100$. It is represented as a $6\times 6$ matrix with $5\times 5$ matrices as coefficients. The corresponding graph is as follows.
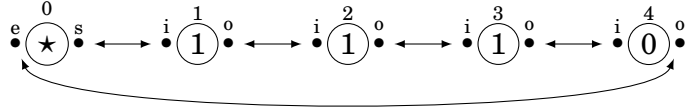
\begin{center}
\begin{tikzpicture}[inner sep=2pt, outer sep=1pt]
	\node[shape=circle,draw] (S) at (0,0) {$\star$};
		\node (SE) at (S.east) {};
		\node (SW) at (S.west) {};
		\node (SN) at (S.north) {};
		\node (So) at (SE.east) {$\bullet$};
		\node (SoL) at (So.north) {\scriptsize{s}};
		\node (Si) at (SW.west) {$\bullet$};
		\node (SiL) at (Si.north) {\scriptsize{e}};
		\node (St) at ($(SN.north)+ (0,.05)$) {\scriptsize{$0$}};
	\node[shape=circle,draw] (A) at (2,0) {$1$};
		\node (AE) at (A.east) {};
		\node (AW) at (A.west) {};
		\node (AN) at (A.north) {};
		\node (Ao) at (AE.east) {$\bullet$};
		\node (AoL) at ($(Ao.north)+ (0,.05)$) {\scriptsize{o}};
		\node (Ai) at (AW.west) {$\bullet$};
		\node (AiL) at ($(Ai.north)+ (0,.05)$) {\scriptsize{i}};
		\node (At) at (AN.north) {\scriptsize{$1$}};
	\node[shape=circle,draw] (B) at (4,0) {$1$};
		\node (BE) at (B.east) {};
		\node (BW) at (B.west) {};
		\node (BN) at (B.north) {};
		\node (Bo) at (BE.east) {$\bullet$};
		\node (BoL) at ($(Bo.north)+ (0,.05)$) {\scriptsize{o}};
		\node (Bi) at (BW.west) {$\bullet$};
		\node (BiL) at ($(Bi.north)+ (0,.05)$) {\scriptsize{i}};
		\node (Bt) at (BN.north) {\scriptsize{$2$}};
	\node[shape=circle,draw] (C) at (6,0) {$1$};
		\node (CE) at (C.east) {};
		\node (CW) at (C.west) {};
		\node (CN) at (C.north) {};
		\node (Co) at (CE.east) {$\bullet$};
		\node (CoL) at ($(Co.north)+ (0,.05)$) {\scriptsize{o}};
		\node (Ci) at (CW.west) {$\bullet$};
		\node (CiL) at ($(Ci.north)+ (0,.05)$) {\scriptsize{i}};
		\node (Ct) at (CN.north) {\scriptsize{$3$}};
	\node[shape=circle,draw] (D) at (8,0) {$0$};
		\node (DE) at (D.east) {};
		\node (DW) at (D.west) {};
		\node (DN) at (D.north) {};
		\node (Do) at (DE.east) {$\bullet$};
		\node (DoL) at ($(Do.north)+ (0,.05)$) {\scriptsize{o}};
		\node (Di) at (DW.west) {$\bullet$};
		\node (DiL) at ($(Di.north)+ (0,.05)$) {\scriptsize{i}};
		\node (Dt) at (DN.north) {\scriptsize{$4$}};
	\draw[<->] (So) to (Ai) {};
	\draw[<->] (Ao) to (Bi) {};
	\draw[<->] (Bo) to (Ci) {};
	\draw[<->] (Co) to (Di) {};
	\draw[<->] (Do) .. controls (8,-1) and (0,-1) .. (Si) {};
\end{tikzpicture}
\end{center}
The corresponding matrix is then:
\[
M_{\star 1110}=
\left(\begin{array}{cccccc}
	0 & 0 & 0 & l_{10} & 0 & 0 \\
	0 & 0 & 0 & 0 & 0 & e_{0}^{\ast}\\
	0 & 0 & 0 & l_{11} & s_{1} & 0 \\
	l_{10}^{\ast} & 0 & l_{11}^{\ast} & 0 & 0 & 0 \\
	0 & 0 & s_{1}^{\ast} & 0 & 0 & 0\\
	0 & e_{0} & 0 & 0 & 0 & 0 \\
\end{array}\right)
\]
where the matrices $l_{10}$, $l_{11}$, $s_{1}$ and $e_{0}$ are defined as:
\[
l_{10} = \left(\begin{array}{ccccc}
	0 & 0 & 0 & 0 & 0\\
	0 & 0 & 0 & 0 & 0\\
	0 & 0 & 0 & 0 & 0\\
	0 & 0 & 0 & 0 & 0\\
	0 & 0 & 0 & 1 & 0
\end{array}\right)
l_{11} = \left(\begin{array}{ccccc}
	0 & 0 & 0 & 0 & 0\\
	0 & 0 & 0 & 0 & 0\\
	0 & 1 & 0 & 0 & 0\\
	0 & 0 & 1 & 0 & 0\\
	0 & 0 & 0 & 0 & 0
\end{array}\right)
\]
\[
s_{0} = \left(\begin{array}{ccccc}
	0 & 0 & 0 & 0 & 0\\
	1 & 0 & 0 & 0 & 0\\
	0 & 0 & 0 & 0 & 0\\
	0 & 0 & 0 & 0 & 0\\
	0 & 0 & 0 & 0 & 0
\end{array}\right)
e_{0} = \left(\begin{array}{ccccc}
	0 & 0 & 0 & 0 & 1\\
	0 & 0 & 0 & 0 & 0\\
	0 & 0 & 0 & 0 & 0\\
	0 & 0 & 0 & 0 & 0\\
	0 & 0 & 0 & 0 & 0
\end{array}\right)
\]

\subsection{The Palindrome DPM}\label{exampledpm}

We will now define a deterministic pointer machine
%with \(2\) pointers
$P=\{Q,\rightarrow\}$ that decides the language of palindromes:
\[\textnormal{Pal}=\{\star a_{0}a_{1}\dots a_{n} \mid \forall i\in\{0,\dots,n\}, a_{i}=a_{n-i}\}\]

To decide \textnormal{Pal}, the machine $P$ will make use of two pointers. It will start with both pointers on $\star$ and move one pointer from left to right over the input and move the second pointer from right to left -- or from the end to the beginning. It will move the pointers alternatively and check if the $i$-th symbol is indeed equal to the $n-i$-th one. It will be defined with five states: an initial state $\textbf{init}$, and states $\textbf{pass}_{1}$, $\textbf{pass}_{2}$, $\textbf{fail}_{1}$ and $\textbf{fail}_{2}$. The subscript will be used to remember which pointer will be moving next. We now define formally the machine $P$.

\begin{definition}[The Palindrome Pointer Machine]
Let \(P = \{Q,\rightarrow\}\) be the DPM($2$) defined as follows: \(Q = \{\textbf{init}, \textbf{pass}_{1}, \textbf{pass}_{2}, \textbf{fail}_{1}, \textbf{fail}_{2}\}\), and the following transitions, where \enquote{\(\cdot\)} (resp.\ \enquote{\(0/1\)}) is a notation standing for any value in \(\{0, 1, \star \}\) (resp.\ in \(\{0, 1\}\)): 
\begin{align}
(\star, \star, \textbf{init}) &\rightarrow (p_{1}+, \epsilon_2, \textbf{pass}_{2}) \label{init-first}\\
(\cdot, \cdot, \textbf{pass}_{2}) &\rightarrow (\epsilon_1, p_{2}-, \textbf{pass}_{1}) \\
(0, 0, \textbf{pass}_{1}) &\rightarrow (p_{1}+, \epsilon_2, \textbf{pass}_{2}) \\
(1, 1, \textbf{pass}_{1}) &\rightarrow (p_{1}+, \epsilon_2, \textbf{pass}_{2}) \\
(0, 1, \textbf{pass}_{1}) &\rightarrow (p_{1}+,\epsilon_2, \textbf{fail}_{2}) \\
(1, 0, \textbf{pass}_{1}) &\rightarrow (p_{1}+,\epsilon_2, \textbf{fail}_{2}) \\
(0/1, 0/1, \textbf{fail}_{1}) &\rightarrow (p_{1}+,\epsilon_2, \textbf{fail}_{2}) \\
(\cdot, \cdot, \textbf{fail}_{2}) &\rightarrow (\epsilon_1, p_{2}-, \textbf{fail}_{1}) \\
(\star, \star, \textbf{pass}_{1}) &\rightarrow \textbf{accept}\label{accept}\\
(\star, \star, \textbf{fail}_{1}) &\rightarrow \textbf{reject}\label{reject}
\end{align}
This transition relation is then completed into a total relation by considering that each configuration not appearing above leads to an acceptation. The machine $P$ is then a \emph{deterministic} pointer machine.
\end{definition}

To represent rejection when encoding this machine as an observation, we will have to create a loop by returning to the initial state (the \enquote{re-initialisation trick}).
The formal encoding defined in the paper works for any machine, but it is a bit heavy.
In the case of \(P\), we use a simple trick to simplify the representation: % as an observation:
%Instead of using the formal encoding defined in the paper -- which works for any machine -- will here use a simple trick -- which works with $P$ but not in general -- to simplify the representation as an observation.
%The trick is as follows: 
we consider the machine defined as above but:
\begin{itemize}
\item without the state $\textbf{init}$ and with initial configuration $(\star,\star,\textbf{fail}_{1})$ -- thus the transition \ref{init-first} is replaced by $(\star, \star, \textbf{fail}_{1}) \rightarrow (p_{1}+, \epsilon_2, \textbf{pass}_{2})$;
\item in which the transition \ref{reject} is replaced by $(\star,\star,\textbf{fail}_{1})\rightarrow(p_{1}+,\textbf{pass}_{2})$.
\end{itemize} This modified machine then either accepts -- if the input is a palindrome -- or loops by going through the initial state over and over -- this loop replaces rejection. By encoding this modified machine as is, we obtain an observation that will accept the same language as $P$; this observation is however simpler than the one we would obtain from applying
mindlessly the translation presented in \cref{encod-non-det}.

\subsection{The Palindrome Observation}

As explained above, we will actually encode the following transitions, and no other -- none of the transitions steps ending with acceptation are encoded since acceptance corresponds to putting a stop to the computation.
\begin{align}
(\star, \star, \textbf{fail}_{1}) &\rightarrow (p_{1}+, \epsilon_2, \textbf{pass}_{2}) \label{init}\\
(\cdot, \cdot, \textbf{pass}_{2}) &\rightarrow (\epsilon_1, p_{2}-, \textbf{pass}_{1}) \label{mov2pass}\\
(\cdot, \cdot, \textbf{fail}_{2}) &\rightarrow (\epsilon_1, p_{2}-, \textbf{fail}_{1}) \label{mov2fail}\\
(0, 0, \textbf{pass}_{1}) &\rightarrow (p_{1}+, \epsilon_2, \textbf{pass}_{2}) \label{00}\\
(1, 1, \textbf{pass}_{1}) &\rightarrow (p_{1}+, \epsilon_2, \textbf{pass}_{2}) \label{11}\\
(0, 1, \textbf{pass}_{1}) &\rightarrow (p_{1}+,\epsilon_2, \textbf{fail}_{2}) \label{01}\\
(1, 0, \textbf{pass}_{1}) &\rightarrow (p_{1}+, \epsilon_2, \textbf{fail}_{2}) \label{10}\\
(0/1, 0/1, \textbf{fail}_{1}) &\rightarrow (p_{1}+, \epsilon_2, \textbf{fail}_{2}) \label{fail}
\end{align}
 
We will moreover use a number of small \enquote{hacks} that will decrease the size of the resulting observation. Indeed, the encoding described in this paper is general and therefore applies blindly to every pointer machines; it may however be optimized in specific cases. The first optimization concerns the set of additional states. The second concerns the way the representation deals with pointers. The third \enquote{hack} will be used to decrease the size of the matrices that represent the pointers \enquote{memory cells}.
 
Here is the first \enquote{hack}. The set of states we are starting from is $Q=\{\textbf{pass}_{1},\textbf{pass}_{2},\textbf{fail}_{1},\textbf{fail}_{2}\}$. As it contains $4$ elements, the set $Q^{\uparrow}$ should have $4+2(4^{2}+2)=40$ elements. Of course, not all these elements are needed here. First, we won't need the specific states introduced to deal with rejection since -- as we already explained -- rejection is already represented by the creation of a loop through the initial state. We are thus left with $4+2\times 4^{2}=36$ states. It turns out that only $4+6=10$ states is enough in this case, as all we need are states to transition:
\begin{itemize}
\item from $\textbf{pass}_{1}$ to $\textbf{pass}_{2}$ (transitions \ref{00} and \ref{11}); we write it as $\textbf{pass}_{1\rightarrow 2}$;
\item from $\textbf{pass}_{1}$ to $\textbf{fail}_{2}$ (transitions \ref{01} and \ref{10}); we write it as $\textbf{error}$;
\item from $\textbf{pass}_{2}$ to $\textbf{pass}_{1}$ (transition \ref{mov2pass}); we write it as $\textbf{pass}_{2\rightarrow 1}$;
\item from $\textbf{fail}_{2}$ to $\textbf{fail}_{1}$ (transition \ref{mov2fail}); we write it as $\textbf{fail}_{2\rightarrow 1}$;
\item from $\textbf{fail}_{1}$ to $\textbf{fail}_{2}$ (transition \ref{fail}); we write it as $\textbf{fail}_{1\rightarrow 2}$;
\item from $\textbf{fail}_{1}$ to $\textbf{pass}_{2}$ (transition \ref{init}); we write it as $\textbf{reinit}$.
\end{itemize}

Let us now explain the second \enquote{hack}. We notice that this machine never uses any \enquote{$\epsilon$ transition} -- i.e.\ the machine moves a pointer at each transition step -- and always alternates the movements of the first and second pointer. The encoding presented in \cref{encod-non-det} makes use of a kind of \enquote{dummy pointer}, and then activates/deactivates pointers in order to move them. We will here get rid of this \enquote{dummy pointer} by considering that it represents the first pointer. As a consequence, the use of the operator $\tau_{0,1}$ will represent the simultaneous activation of a pointer and the deactivation of the other pointer. Moreover, it has an effect on the set of additional states needed to encode the operators: since we simultaneously activate a pointer and deactivate the other, we can get rid of the six additional states explained above and work with the $4$ states of the original pointer machine. This will however complicate the understanding of the encoding in that an operator encoding a given transition will also encode the \enquote{recording} of the value read during the previous transition into memory cells. i.e.\ we will consider operators that are a mixing of the operators $\text{bf}^{c}_{j,\textbf{q'}}+\text{ff}^{c}_{j,\textbf{q'}}$ for the currently encoded transition and the operator $\text{rec}_{k,\textbf{q}}^{c'}$ encoding the recording of the value read during the previous transition. This complication is however worth considering since it will greatly reduce the size of the resulting matrices. 

%\textbf{ALT} \textcolor{gray}{Let us now explain the second \enquote{hack}.
%One may remark that this machine alternatively moves the first and the second pointer at each transition.
%The encoding presented in this paper makes use of a kind of \enquote{dummy pointer} that is at each transition activated and desactivated to leave the room for the next \enquote{main} pointer.
%This \enquote{dummy pointer} will be here considered to be the first pointer, and the operator $\tau_{0,1}$ will represent the simultaneous activation of a pointer and the deactivation of the other.
%No \enquote{buffer} between the desactivation of a pointer and the activation of another is needed, so that we can get rid of the six additional states explained above and work with the $4$ states of the original pointer machine.
%This complicates the understanding of the encoding, because a given transition is encoded by an operator that simultaneously encode this transition (the role devoted to the $\text{bf}^{c}_{j,\textbf{q'}}+\text{ff}^{c}_{j,\textbf{q'}}$ part) and the \enquote{recording} of the value read during the previous transition into memory cells (the $\text{rec}_{k,\textbf{q}}^{c'}$ part).
%However, this simplifies the encoding as the size of the resulting matrices decreases.}

Lastly, let us recall the value last read by a pointer is stored as a state in the encoding of machines. In the general encoding presented in this paper, we store those as $6\times 6$ matrices, which is coherent with the fact that the integer answers belong to a six-elements set $\{0i,0o,1i,1o,s,e\}$. However, all we will need is the information about the symbol read, and we won't care whether we read this symbol by moving left or moving right. This allows us to use $3\times 3$ matrices as memory cells, decreasing once again the size of the resulting observation.

Finally, each transition will be represented as an element of $\vn{M}_{6}(\complexN)\otimes\vn{G}\otimes\vn{M}_{3}(\complexN)\otimes\vn{M}_{3}(\complexN)\otimes\vn{M}_{4}(\complexN)$. The algebra $\vn{M}_{6}(\complexN)$ will allow the observation to interact with the integer, i.e.\ ask what is the next or previous symbol on the input tape. The two copies of $\vn{M}_{3}(\complexN)$ represent the \enquote{memory cells} corresponding to the pointers, i.e.\ they will be used to record the last values read by the pointers. The algebra $\vn{M}_{4}(\complexN)$ is used to deal with states. We represent the $6\times 6$, $3\times 3$ matrices and $4\times 4$ matrices using the respective bases $\{0i,0o,1i,1o,s,e\}$, $\{0,1,\star\}$ and $\{\textbf{pass}_{1},\textbf{pass}_{2},\textbf{fail}_{1},\textbf{fail}_{2}\}$ with basis elements considered in this order exactly.

We first represent the transitions \ref{mov2pass} and \ref{mov2fail}. We represent them simultaneously to gain some space; this is possible because they are the same transition but for the involved states. These transitions do not depend on the values read by the pointers, but will write down the new values in the two copies of $\vn{M}_{3}(\complexN)$ that represent the \enquote{memory cells}. It will therefore be represented as a sum of the following nine operators:
\[
\left(\begin{array}{cccccc}
	0 & 0 & 0 & 0 & 0 & 0\\
	0 & 1 & 0 & 0 & 0 & 0\\
	0 & 0 & 0 & 0 & 0 & 0\\
	0 & 0 & 0 & 0 & 0 & 0\\
	0 & 0 & 0 & 0 & 0 & 0\\
	0 & 0 & 0 & 0 & 0 & 0
\end{array}\right)
\otimes
\tau_{0,1}
\otimes
\left(\begin{array}{ccc}
	1 & 1 & 1\\
	0 & 0 & 0\\
	0 & 0 & 0
\end{array}\right)
\otimes
\left(\begin{array}{ccc}
	1 & 0 & 0\\
	0 & 0 & 0\\
	0 & 0 & 0
\end{array}\right)
\otimes
\left(\begin{array}{cccc}
	0 & 1 & 0 & 0 \\
	0 & 0 & 0 & 0 \\
	0 & 0 & 0 & 1 \\
	0 & 0 & 0 & 0 
\end{array}\right)
\]
\[
\left(\begin{array}{cccccc}
	0 & 0 & 0 & 0 & 0 & 0\\
	0 & 0 & 0 & 0 & 0 & 0\\
	0 & 0 & 0 & 0 & 0 & 0\\
	0 & 1 & 0 & 0 & 0 & 0\\
	0 & 0 & 0 & 0 & 0 & 0\\
	0 & 0 & 0 & 0 & 0 & 0
\end{array}\right)
\otimes
\tau_{0,1}
\otimes
\left(\begin{array}{ccc}
	1 & 1 & 1\\
	0 & 0 & 0\\
	0 & 0 & 0
\end{array}\right)
\otimes
\left(\begin{array}{ccc}
	0 & 0 & 0\\
	0 & 0 & 0\\
	0 & 0 & 1
\end{array}\right)
\otimes
\left(\begin{array}{cccc}
	0 & 1 & 0 & 0 \\
	0 & 0 & 0 & 0 \\
	0 & 0 & 0 & 1 \\
	0 & 0 & 0 & 0 
\end{array}\right)
\]
\[
\left(\begin{array}{cccccc}
	0 & 0 & 0 & 0 & 0 & 0\\
	0 & 0 & 0 & 0 & 0 & 0\\
	0 & 0 & 0 & 0 & 0 & 0\\
	0 & 0 & 0 & 0 & 0 & 0\\
	0 & 1 & 0 & 0 & 0 & 0\\
	0 & 0 & 0 & 0 & 0 & 0
\end{array}\right)
\otimes
\tau_{0,1}
\otimes
\left(\begin{array}{ccc}
	1 & 1 & 1\\
	0 & 0 & 0\\
	0 & 0 & 0
\end{array}\right)
\otimes
\left(\begin{array}{ccc}
	1 & 0 & 0\\
	0 & 0 & 0\\
	0 & 0 & 0
\end{array}\right)
\otimes
\left(\begin{array}{cccc}
	0 & 1 & 0 & 0 \\
	0 & 0 & 0 & 0 \\
	0 & 0 & 0 & 1 \\
	0 & 0 & 0 & 0 
\end{array}\right)
\]
\[
\left(\begin{array}{cccccc}
	0 & 0 & 0 & 0 & 0 & 0\\
	0 & 0 & 0 & 1 & 0 & 0\\
	0 & 0 & 0 & 0 & 0 & 0\\
	0 & 0 & 0 & 0 & 0 & 0\\
	0 & 0 & 0 & 0 & 0 & 0\\
	0 & 0 & 0 & 0 & 0 & 0
\end{array}\right)
\otimes
\tau_{0,1}
\otimes
\left(\begin{array}{ccc}
	0 & 0 & 0\\
	1 & 1 & 1\\
	0 & 0 & 0
\end{array}\right)
\otimes
\left(\begin{array}{ccc}
	1 & 0 & 0\\
	0 & 0 & 0\\
	0 & 0 & 0
\end{array}\right)
\otimes
\left(\begin{array}{cccc}
	0 & 1 & 0 & 0 \\
	0 & 0 & 0 & 0 \\
	0 & 0 & 0 & 1 \\
	0 & 0 & 0 & 0 
\end{array}\right)
\]
\[
\left(\begin{array}{cccccc}
	0 & 0 & 0 & 0 & 0 & 0\\
	0 & 0 & 0 & 0 & 0 & 0\\
	0 & 0 & 0 & 0 & 0 & 0\\
	0 & 0 & 0 & 1 & 0 & 0\\
	0 & 0 & 0 & 0 & 0 & 0\\
	0 & 0 & 0 & 0 & 0 & 0
\end{array}\right)
\otimes
\tau_{0,1}
\otimes
\left(\begin{array}{ccc}
	0 & 0 & 0\\
	1 & 1 & 1\\
	0 & 0 & 0
\end{array}\right)
\otimes
\left(\begin{array}{ccc}
	0 & 0 & 0\\
	0 & 0 & 0\\
	0 & 0 & 1
\end{array}\right)
\otimes
\left(\begin{array}{cccc}
	0 & 1 & 0 & 0 \\
	0 & 0 & 0 & 0 \\
	0 & 0 & 0 & 1 \\
	0 & 0 & 0 & 0 
\end{array}\right)
\]
\[
\left(\begin{array}{cccccc}
	0 & 0 & 0 & 0 & 0 & 0\\
	0 & 0 & 0 & 0 & 0 & 0\\
	0 & 0 & 0 & 0 & 0 & 0\\
	0 & 0 & 0 & 0 & 0 & 0\\
	0 & 0 & 0 & 1 & 0 & 0\\
	0 & 0 & 0 & 0 & 0 & 0
\end{array}\right)
\otimes
\tau_{0,1}
\otimes
\left(\begin{array}{ccc}
	0 & 0 & 0\\
	1 & 1 & 1\\
	0 & 0 & 0
\end{array}\right)
\otimes
\left(\begin{array}{ccc}
	1 & 0 & 0\\
	0 & 0 & 0\\
	0 & 0 & 0
\end{array}\right)
\otimes
\left(\begin{array}{cccc}
	0 & 1 & 0 & 0 \\
	0 & 0 & 0 & 0 \\
	0 & 0 & 0 & 1 \\
	0 & 0 & 0 & 0 
\end{array}\right)
\]
\[
\left(\begin{array}{cccccc}
	0 & 0 & 0 & 0 & 0 & 0\\
	0 & 0 & 0 & 0 & 0 & 1\\
	0 & 0 & 0 & 0 & 0 & 0\\
	0 & 0 & 0 & 0 & 0 & 0\\
	0 & 0 & 0 & 0 & 0 & 0\\
	0 & 0 & 0 & 0 & 0 & 0
\end{array}\right)
\otimes
\tau_{0,1}
\otimes
\left(\begin{array}{ccc}
	0 & 0 & 0\\
	0 & 0 & 0\\
	1 & 1 & 1
\end{array}\right)
\otimes
\left(\begin{array}{ccc}
	1 & 0 & 0\\
	0 & 0 & 0\\
	0 & 0 & 0
\end{array}\right)
\otimes
\left(\begin{array}{cccc}
	0 & 1 & 0 & 0 \\
	0 & 0 & 0 & 0 \\
	0 & 0 & 0 & 1 \\
	0 & 0 & 0 & 0 
\end{array}\right)
\]
\[
\left(\begin{array}{cccccc}
	0 & 0 & 0 & 0 & 0 & 0\\
	0 & 0 & 0 & 0 & 0 & 0\\
	0 & 0 & 0 & 0 & 0 & 0\\
	0 & 0 & 0 & 0 & 0 & 1\\
	0 & 0 & 0 & 0 & 0 & 0\\
	0 & 0 & 0 & 0 & 0 & 0
\end{array}\right)
\otimes
\tau_{0,1}
\otimes
\left(\begin{array}{ccc}
	0 & 0 & 0\\
	0 & 0 & 0\\
	1 & 1 & 1
\end{array}\right)
\otimes
\left(\begin{array}{ccc}
	0 & 0 & 0\\
	0 & 0 & 0\\
	0 & 0 & 1
\end{array}\right)
\otimes
\left(\begin{array}{cccc}
	0 & 1 & 0 & 0 \\
	0 & 0 & 0 & 0 \\
	0 & 0 & 0 & 1 \\
	0 & 0 & 0 & 0 
\end{array}\right)
\]
\[
\left(\begin{array}{cccccc}
	0 & 0 & 0 & 0 & 0 & 0\\
	0 & 0 & 0 & 0 & 0 & 0\\
	0 & 0 & 0 & 0 & 0 & 0\\
	0 & 0 & 0 & 0 & 0 & 0\\
	0 & 0 & 0 & 0 & 0 & 1\\
	0 & 0 & 0 & 0 & 0 & 0
\end{array}\right)
\otimes
\tau_{0,1}
\otimes
\left(\begin{array}{ccc}
	0 & 0 & 0\\
	0 & 0 & 0\\
	1 & 1 & 1
\end{array}\right)
\otimes
\left(\begin{array}{ccc}
	1 & 0 & 0\\
	0 & 0 & 0\\
	0 & 0 & 0
\end{array}\right)
\otimes
\left(\begin{array}{cccc}
	0 & 1 & 0 & 0 \\
	0 & 0 & 0 & 0 \\
	0 & 0 & 0 & 1 \\
	0 & 0 & 0 & 0 
\end{array}\right)
\]

Notice that these operators are all morally the same. One should consider them by groups of three. The first three correspond to the case when the value read by the first pointer during the last transition was a $0$. In this case, the integer answered on the second column of the $\vn{M}_{6}(\complexN)$ matrix since this column corresponds to the basis element $0o$ (for \enquote{0 out}) and the first pointer moves forward. Then the first operator corresponds to the case of the last value read by the second pointer is also a $0$. Thus, to ask what is the next value read by this second pointer one needs to activate it -- this is the role of $\tau_{0,1}$ -- and the machine needs to output on the second row of the $6\times 6$ matrix: this is because we want to know what symbol \emph{preceded} the last $0$ read by the second pointer (remember that the second pointer moves backwards). The second operator in this first group corresponds to the case when the last value read by the second pointer was $1$. Then, the $6\times 6$ matrix used moves from the basis element $0o$ -- which corresponds to the previous answer given by the integer -- to the basis element $1o$ -- to ask what symbol preceded the last $1$ read by the second pointer. The third operator corresponds to the case when the last symbol read by the second pointer was a $\star$. Finally, the two following groups of three matrices play the same role as this first group in the eventuality that the last value read by the first pointer was a $1$ or a $\star$.

Notice also the right-hand matrix in each of these transitions. This matrix takes care of the states. This is why there are two non-zero coefficients in those matrices here: one represents the transition \ref{mov2pass} and the second represent the transition \ref{mov2fail}.

Let us now deal with the representation of the transition \ref{00}. It records the last value read by the second pointer, which should be $0$, and asks the value following the last value read by the first pointer, which is $0$ also. Thus, the first matrix goes from $0i$ (since the second pointer goes backwards) to $0i$ (since the first pointer moves forward). Then, the first $3\times 3$ matrix ensures that the last value read by the first pointer was indeed $0$, and the second $3\times 3$ matrix records the new value read by the second pointer: whichever value was stored beforehand, it is now replaced by a $0$. Lastly, the $4\times 4$ matrix deals with the change of states. As before, we use the fact that this transition is almost similar to the transition \ref{fail} when both values are equal to $0$ and represent both by a single operator: this is shown by the fact that the right-hand matrix not only contains a transition from $\textbf{pass}_{1}$ to $\textbf{pass}_{2}$ (for transition \ref{00}) but also a transition from $\textbf{fail}_{1}$ to $\textbf{fail}_{2}$.
\[
\left(\begin{array}{cccccc}
	1 & 0 & 0 & 0 & 0 & 0\\
	0 & 0 & 0 & 0 & 0 & 0\\
	0 & 0 & 0 & 0 & 0 & 0\\
	0 & 0 & 0 & 0 & 0 & 0\\
	0 & 0 & 0 & 0 & 0 & 0\\
	0 & 0 & 0 & 0 & 0 & 0
\end{array}\right)
\otimes
\tau_{0,1}
\otimes
\left(\begin{array}{ccc}
	1 & 0 & 0\\
	0 & 0 & 0\\
	0 & 0 & 0
\end{array}\right)
\otimes
\left(\begin{array}{ccc}
	1 & 1 & 1\\
	0 & 0 & 0\\
	0 & 0 & 0
\end{array}\right)
\otimes
\left(\begin{array}{cccc}
	0 & 0 & 0 & 0 \\
	1 & 0 & 0 & 0 \\
	0 & 0 & 0 & 0 \\
	0 & 0 & 1 & 0 
\end{array}\right)
\]

The operator encoding the transitions \ref{11}, \ref{01} and \ref{10} and the corresponding cases of transition \ref{fail} are quite similar and are shown below in that order.
\[
\left(\begin{array}{cccccc}
	0 & 0 & 0 & 0 & 0 & 0\\
	0 & 0 & 0 & 0 & 0 & 0\\
	0 & 0 & 1 & 0 & 0 & 0\\
	0 & 0 & 0 & 0 & 0 & 0\\
	0 & 0 & 0 & 0 & 0 & 0\\
	0 & 0 & 0 & 0 & 0 & 0
\end{array}\right)
\otimes
\tau_{0,1}
\otimes
\left(\begin{array}{ccc}
	0 & 0 & 0\\
	0 & 1 & 0\\
	0 & 0 & 0
\end{array}\right)
\otimes
\left(\begin{array}{ccc}
	0 & 0 & 0\\
	1 & 1 & 1\\
	0 & 0 & 0
\end{array}\right)
\otimes
\left(\begin{array}{cccc}
	0 & 0 & 0 & 0 \\
	1 & 0 & 0 & 0 \\
	0 & 0 & 0 & 0 \\
	0 & 0 & 1 & 0 
\end{array}\right)
\]
\[
\left(\begin{array}{cccccc}
	0 & 0 & 1 & 0 & 0 & 0\\
	0 & 0 & 0 & 0 & 0 & 0\\
	0 & 0 & 0 & 0 & 0 & 0\\
	0 & 0 & 0 & 0 & 0 & 0\\
	0 & 0 & 0 & 0 & 0 & 0\\
	0 & 0 & 0 & 0 & 0 & 0
\end{array}\right)
\otimes
\tau_{0,1}
\otimes
\left(\begin{array}{ccc}
	1 & 0 & 0\\
	0 & 0 & 0\\
	0 & 0 & 0
\end{array}\right)
\otimes
\left(\begin{array}{ccc}
	0 & 0 & 0\\
	1 & 1 & 1\\
	0 & 0 & 0
\end{array}\right)
\otimes
\left(\begin{array}{cccc}
	0 & 0 & 0 & 0 \\
	0 & 0 & 0 & 0 \\
	0 & 0 & 0 & 0 \\
	1 & 0 & 1 & 0 
\end{array}\right)
\]
\[
\left(\begin{array}{cccccc}
	0 & 0 & 0 & 0 & 0 & 0\\
	0 & 0 & 0 & 0 & 0 & 0\\
	1 & 0 & 0 & 0 & 0 & 0\\
	0 & 0 & 0 & 0 & 0 & 0\\
	0 & 0 & 0 & 0 & 0 & 0\\
	0 & 0 & 0 & 0 & 0 & 0
\end{array}\right)
\otimes
\tau_{0,1}
\otimes
\left(\begin{array}{ccc}
	0 & 0 & 0\\
	0 & 1 & 0\\
	0 & 0 & 0
\end{array}\right)
\otimes
\left(\begin{array}{ccc}
	1 & 1 & 1\\
	0 & 0 & 0\\
	0 & 0 & 0
\end{array}\right)
\otimes
\left(\begin{array}{cccc}
	0 & 0 & 0 & 0 \\
	0 & 0 & 0 & 0 \\
	0 & 0 & 0 & 0 \\
	1 & 0 & 1 & 0 
\end{array}\right)
\]

Finally, we represent the transition \ref{init}. It is represented as the following operator:
\[
\left(\begin{array}{cccccc}
	0 & 0 & 0 & 0 & 0 & 0\\
	0 & 0 & 0 & 0 & 0 & 0\\
	0 & 0 & 0 & 0 & 0 & 0\\
	0 & 0 & 0 & 0 & 0 & 0\\
	0 & 0 & 0 & 0 & 1 & 0\\
	0 & 0 & 0 & 0 & 0 & 0
\end{array}\right)
\otimes
\tau_{0,1}
\otimes
\left(\begin{array}{ccc}
	0 & 0 & 0\\
	0 & 0 & 0\\
	0 & 0 & 1
\end{array}\right)
\otimes
\left(\begin{array}{ccc}
	0 & 0 & 0\\
	0 & 0 & 0\\
	1 & 1 & 1
\end{array}\right)
\otimes
\left(\begin{array}{cccc}
	0 & 0 & 0 & 0 \\
	0 & 0 & 1 & 0 \\
	0 & 0 & 0 & 0 \\
	0 & 0 & 0 & 0 
\end{array}\right)
\]

\end{document}